\documentclass[iop]{emulateapj}

\usepackage{apjfonts}
\usepackage{graphicx}
\usepackage{epstopdf}
\usepackage{ulem}
\usepackage[utf8]{inputenc}
\usepackage[T1]{fontenc}
 \usepackage{url}
\usepackage{hyperref}

\makeatletter
\def\url@leostyle{%
 \@ifundefined{selectfont}{\def\UrlFont{\sf}}{\def\UrlFont{\small\ttfamily}}}
\makeatother
\begin{document}

\newcommand{\ls}{{_<\atop^{\sim}}}
\newcommand{\gs}{{_>\atop^{\sim}}}
\def \spose#1{\hbox  to 0pt{#1\hss}}  
\def \ls{\mathrel{\spose{\lower 3pt\hbox{$\sim$}}\raise  2.0pt\hbox{$<$}}}
\def \gs{\mathrel{\spose{\lower  3pt\hbox{$\sim$}}\raise 2.0pt\hbox{$>$}}}
\newcommand{\Ha}{\hbox{{\rm H}$\alpha$}}
\newcommand{\Hb}{\hbox{{\rm H}$\beta$}}
\newcommand{\OIII}{\hbox{[{\rm O}\kern 0.1em{\sc iii}]}}
\newcommand{\NII}{\hbox{[{\rm N}\kern 0.1em{\sc ii}]}}


\title{The Dependence of Quenching upon the Inner Structure of Galaxies at $0.5 \le \lowercase{z} < 0.8$ in the DEEP2/AEGIS Survey}

\shorttitle{The Dependence of Quenching upon Inner Galactic Structure}
\shortauthors{Cheung et al.}


\author{Edmond Cheung\altaffilmark{1}, S. M. Faber\altaffilmark{1} \altaffilmark{2}, David C. Koo\altaffilmark{1} \altaffilmark{2}, Aaron A. Dutton\altaffilmark{3},  Luc Simard\altaffilmark{4},  Elizabeth J. McGrath\altaffilmark{1},  J.-S. Huang\altaffilmark{5}, Eric F. Bell\altaffilmark{18}, Avishai Dekel\altaffilmark{16}, Jerome J. Fang\altaffilmark{1}, Samir Salim\altaffilmark{7}, G. Barro\altaffilmark{1}, K. Bundy\altaffilmark{11}, A. L. Coil\altaffilmark{10}, Michael C. Cooper\altaffilmark{14} \altaffilmark{\dag},C.J. Conselice\altaffilmark{12}, M. Davis\altaffilmark{21}, A. Dom\'{i}nguez\altaffilmark{13},  Susan A. Kassin\altaffilmark{17} \altaffilmark{22}, Dale D. Kocevski\altaffilmark{1}, Anton M. Koekemoer\altaffilmark{8}, Lihwai Lin\altaffilmark{20}, Jennifer M. Lotz\altaffilmark{7}, J. A. Newman\altaffilmark{9}, Andrew C. Phillips\altaffilmark{1}, D. J. Rosario\altaffilmark{15}, Benjamin J. Weiner\altaffilmark{19}, C. N. A. Willmer\altaffilmark{6}}

\email{ec2250@gmail.com}

\altaffiltext{1}{Department of Astronomy and Astrophysics, 1156 High Street, University of California, Santa Cruz, CA 95064}
\altaffiltext{2}{UCO/Lick Observatory; University of California, 1156 High Street, Santa Cruz, CA 95064}
\altaffiltext{3}{Department of Physics and Astronomy, Elliot Building, 3800 Finnerty Road, University of Victoria, Victoria BC, V8P 5C2, Canada}
\altaffiltext{4}{Herzberg Institute of Astrophysics, National Research Council of Canada, 5071 West Saanich Road, Victoria, BC, V9E 2E7, Canada}
\altaffiltext{5}{Harvard-Smithsonian Center for Astrophysics, 60 Garden State, Cambridge, MA 02138, USA}
\altaffiltext{6}{Steward Observatory, University of Arizona, 933 North Cherry Avenue, Tucson, AZ 85721, USA}
\altaffiltext{7}{National Optical Astronomical Observatories, 950 N. Cherry Avenue, Tucson, AZ 85719, USA}
\altaffiltext{8}{Space Telescope Science Institute, 3700 San Martin Drive, Baltimore, MD 21218}
\altaffiltext{9}{Department of Physics and Astronomy, University of Pittsburgh, 3941 O\textquoteright Hara Street, Pittsburgh, PA 15260, USA}
\altaffiltext{10}{Department of Physics, Center for Astrophysics and Space Sciences, University of California at San Diego, 9500 Gilman Dr., La Jolla, San Diego, CA 92093}
\altaffiltext{11}{Kavli Institute for the Physics and Mathematics of the Universe, University of Tokyo, Kashiwa, 277-8582, Japan}
\altaffiltext{12}{University of Nottingham, School of Physics and Astronomy, Nottingham NG7 2RD, UK}
\altaffiltext{13}{Department of Physics and Astronomy, University of California, Riverside, CA 92521, USA}
\altaffiltext{14}{Center for Galaxy Evolution, Department of Physics and Astronomy, University of California, Irvine, 4129 Frederick Reines Hall, Irvine, CA 92697, USA}
\altaffiltext{15}{Max Planck Institute for extraterrestrial Physics, PO Box 1312, Giessenbachstrasse, 85741 Garching, Germany}
\altaffiltext{16}{Racah Institute of Physics, The Hebrew University, Jerusalem 91904, Israel}
\altaffiltext{17}{Astrophysics Science Division, Goddard Space Flight Center, Code 665, Greenbelt, MD 20771}
\altaffiltext{18}{Department of Astronomy, University of Michigan, 500 Church St., Ann Arbor, MI 48109, USA}
\altaffiltext{19}{Steward Observatory, 933 N. Cherry St., University of Arizona, Tucson, AZ 85721, USA}
\altaffiltext{20}{Institute of Astronomy and Astrophysics, Academia Sinica, Taipei 106,Taiwan}
\altaffiltext{21}{Department of Astronomy, University of California, Berkeley, CA 94720}
\altaffiltext{22}{NASA Postdoctoral Program Fellow}
\altaffiltext{\dag}{Hubble Fellow}

\slugcomment{Accepted for publication in ApJ; 2012 October 11}


\begin{abstract}

The shutdown of star formation in galaxies is generally termed `quenching'. Quenching may occur through a variety of processes, e.g., AGN feedback, stellar feedback, or the shock heating of gas in the dark matter halo. However, which mechanism(s) is, in fact, responsible for quenching is still in question. This paper addresses quenching by searching for traces of possible quenching processes through their effects on galaxy structural parameters such as stellar mass ($M_*$), $M_*/r_{\rm e}$, surface stellar mass density ($\sim M_*/r_{\rm e}^2$), and S\'ersic index ($n$). We analyze the rest-frame $U-B$ color correlations versus these structural parameters using a sample of galaxies in the redshift range $0.5\le z<0.8$ from the DEEP2/AEGIS survey. In addition to global radii, stellar masses, and S\'ersic parameters, we also use `bulge' and `disk' photometric measurements from GIM2D fits to $HST$/ACS $V$ and $I$ images. We assess the tightness of the color relationships by measuring their `overlap regions', defined as the area in color-parameter space in which red and blue galaxies overlap; the parameter that minimizes these overlap regions is considered to be the most effective color discriminator. We find that S\'ersic index ($n$) has the smallest overlap region among all tested parameters and resembles a step-function with a threshold value of $n=2.3$. There exists, however, a significant population of outliers with blue colors yet high $n$ values that seem to contradict this behavior; they make up $\approx40\%$ of $n>2.3$ galaxies. We hypothesize that their S\'ersic values may be distorted by bursts of star formation, AGNs, and/or poor fits, leading us to consider central surface {\it stellar mass density}, $\Sigma_{\rm 1 kpc}^*$, as an alternative to S\'ersic index. Not only does $\Sigma_{\rm 1 kpc}^*$ correct the outliers, it also forms a tight relationship with color, suggesting that the {\it innermost structure of galaxies is most physically linked with quenching}. Furthermore, at $z\sim0.65$, the majority of the blue cloud galaxies cannot simply fade onto the red sequence since their GIM2D bulge masses are only half as large on average as the bulge masses of similar red sequence galaxies, thus demonstrating that stellar mass must absolutely increase at the centers of galaxies as they quench. We discuss a two-stage model for quenching in which galaxy star formation rates are controlled by their dark halos while they are still in the blue cloud and a second quenching process sets in later, associated with the central stellar mass build-up. The mass build-up is naturally explained by any non-axisymmetric features in the potential, such as those induced by mergers and/or disk instabilities. However, the identity of the second quenching agent is still unknown. We have placed our data catalog on line.

\end{abstract}

\keywords{galaxies: bulges --- galaxies: formation --- galaxies:
  evolution --- galaxies: structure --- galaxies: fundamental
  parameters}


\section{Introduction} \label{sec:introduction}

With the advent of large galaxy surveys, the color bimodality of the galaxy population has become well-characterized \citep{lin99, strateva01, im02, blanton03, kauffmann03, bell04}. Galaxy counts back in time revealed that the number of red galaxies has at least doubled since $z\sim1$ while the number of blue galaxies has remained relatively constant \citep{bell04, bundy06, faber07, arnouts07, brown07, ilbert10, dominguez11, goncalves12}. A natural interpretation is that galaxies evolve from blue to red with time, i.e., from star-forming to `quenched'. Later measurements of star formation rates confirmed that blue galaxies create stars at a high rate while red galaxies show little to no star formation \citep{salim05, salim07, noeske07, zheng07}. Moreover, star formation rates in blue galaxies correlate well with stellar mass, forming the `Main Sequence' of star formation. In non-dusty red galaxies, however, star formation is generally much lower than that of blue ones \citep{salim05}. This abrupt jump in star formation rate across colors motivates the search for a quenching process. For simplicity, we define quenching to be a process that permanently turns a blue star-forming galaxy into a red non-star-forming one.\footnote{Rejuvenation of star formation in quiescent spheroids through gas and/or satellite infall has been proposed to explain the observed blue spheroids seen in various works \citep[e.g.,][]{kannappan09, schawinski09}. In this paper, we do not consider this process.}
 
Many quenching mechanisms have been proposed, but they can generally be categorized into two classes. The first class is internal processes; these act to either expel the gas already in a galaxy or render it inert to star formation. Examples of internal processes include feedback from starbursts and active galactic nuclei (AGN), both of which may be triggered by mergers. They act to heat the surrounding gas and/or drive winds out of the galaxy \citep[e.g.,][]{sanders88, springel05, murray05, cox08, ciotti09, alexander10}. Another example of an internal process is morphological quenching \citep{martig09}. In this model, the presence of a dominant bulge stabilizes the gaseous disk against gravitational instabilities needed for star formation. 

The second class contains external processes, which we define as acting to prevent gas from accreting onto a galaxy in the first place. The main external process is halo mass quenching \citep{silk77, rees&ostriker77, blumenthal84, birnboim03, keres05, dekel06, cattaneo06}; this posits that dark matter halos above a critical halo mass establish virial shocks that stop the flow of cold gas onto their central galaxies. Additional examples are AGN `radio mode' feedback \citep{croton06} and gravitational heating \citep{khochfar08, birnboim10}, both of which can be considered as variants of halo mass quenching since both mechanisms require massive halos. 

According to our definition, mergers do not qualify as an external process since they act to exhaust and/or remove existing gas already within a galaxy. By the same token, ram pressure stripping \citep{gunn72} is also not considered an external process since it strips gas from a galaxy. Furthermore, this paper only concentrates on quenching processes that affect the central galaxy of a halo. According to \cite{gerke05}, who used a sample of DEEP2 galaxies ($\sim25\%$ of the total DEEP2 sample), only $\sim32\%$ of DEEP2 galaxies are in groups, meaning $\sim68\%$ of these galaxies are in the field. These field galaxies would be centrals and additionally, since each group contains one central, the percentage of centrals in this DEEP2 sample is at least $\sim68\%$. Assuming this sample is representative of the entire DEEP2 dataset, we can conclude that most of our galaxies are centrals. Thus we will not consider mechanisms that affect satellites, i.e., strangulation and harassment \citep{larson80, moore96}.

These quenching processes may imprint themselves on the structure of a galaxy, e.g., major majors can create highly concentrated galaxies. The prospect of detecting quenching mechanisms at work via observable changes in structural parameters has motivated many previous works. One of the first parameters explored was luminosity. Using an early SDSS sample, \cite{strateva01} found that galaxies are bimodal in color, i.e., galaxies generally lie within the red sequence or the blue cloud. However, while galaxies are well separated in color, they overlap over almost the entire range of luminosity, indicating that luminosity is not the main driver of galaxy color. 

Later, stellar mass was explored; hereafter, mass refers to stellar mass unless otherwise stated. For a sample of local SDSS galaxies, \cite{kauffmann03} found that the correlations between the star formation history indicators $\rm{D}_{n}(4000)$ and H$\delta_A$ (which can also be thought of as a proxy for galaxy color) and mass are significantly better than that of the $g$-band luminosity. They further found that galaxies divide into two distinct families at a stellar mass of $3 \times 10^{10} ~\rm M_{\odot}$.

Recently, additional structural parameters have been introduced. Using an SDSS sample, \cite{kauffmann06} found that the galaxy surface mass density ($\sim M_*/r_{\rm e}^2$) produced an even sharper division in specific star formation rate (SSFR) than stellar mass (see also \citealp{brinchmann04, maier09}). They suggested that high surface stellar mass density is connected to the creation of a bulge and the quenching of a galaxy.

\cite{franx08} intercompared several of the aforementioned color-parameter correlations in the redshift range $0< z <3$ using data from the FIREWORKS catalog \citep{wuyts08}. Confirming \cite{kauffmann06}'s result, they showed that surface mass density better separates red and blue galaxies than stellar mass alone. \cite{franx08} also examined a second structural parameter, the ``inferred velocity dispersion'' ($\sim M_*/r_{\rm e}$), and found that the inferred velocity dispersion also better distinguishes red and blue galaxies than mass. 

Besides these structural parameters, S\'ersic index ($n$) has also been explored. \cite{driver06} and \cite{allen06} observed a clear bimodal distribution in both the rest-frame $u-r$ color and $n$ in the Millennium Galaxy Catalog. A similar trend with SDSS galaxies was seen by \cite{blanton03} and \cite{schiminovich07}. \cite{bell08} showed that $n$ is an even better color discriminator than surface mass density. However, several outliers were noted, and he concluded that high $n$ is a necessary (but not sufficient) condition for quiescence. Recently, \cite{wuyts11} and \cite{bell12} found that the correlation between quenching and $n$ was in place since at least $z\sim2.5$. 

A study by \cite{mendez11} supports the implications of the relationship between S\'ersic index and quiescence. Using a sample of DEEP2/AEGIS galaxies at $0.4<z<1.2$, they compared the morphological parameters (CAS, G/$M_{20}$, and $B/T$) of galaxies in the green valley -- galaxies with colors that lie between the blue and red peak in the color bimodality -- to those in the blue cloud and red sequence. They found that most green valley galaxies are still disks but are building up their central bulge, in that they have higher concentrations and higher $B/T$ ratios than blue galaxies and less than red galaxies. In other words, they found that the bulges of galaxies are being created or augmented in the evolution of a galaxy from the blue cloud, through the green valley, and finally onto the red sequence.

A recent study by \cite{wake12} adds SDSS central velocity dispersion to the list of previously considered structural parameters.  It is also the first study to compare the efficacy of S\'ersic indices head to head versus other variables.  They find that central velocity dispersion leaves the weakest residual color trends with other parameters and conclude quenching correlates most strongly with central velocity dispersion.

The dependence of quiescence on halo properties has thus far been measured only {\it statistically}, by looking at the probability that a galaxy is quenched as a function of some mass and/or surrounding density.  \cite{peng10} found that just two processes -- ``stellar mass quenching'', which correlates directly with galaxy stellar mass, and ``environmental quenching'', which correlates directly with local environmental density -- can accurately describe the quenching probabilities of SDSS galaxies. A later paper \citep{peng11} divided centrals from satellites and found that central quenching -- of relevance here -- had no environment dependence but related only to stellar mass. A similar study by \cite{woo12} introduced halo mass, which \cite{peng10} had not considered, and found that central quenching correlated better with halo mass than with stellar mass.  However, it is important to note that, regardless of whether halo mass is better than stellar mass, it is clearly {\it not} as predictive as structural variables such as S\'ersic index or central velocity dispersion.  We expound on this statement in the discussion of this paper, but a cursory examination of the SSFR as a function of halo mass from \cite{conroy09} (Fig.  8) shows that star formation only gradually changes as a function of halo mass. Whereas the plots of color as a function of S\'ersic index and central velocity dispersion from \cite{wake12} (Fig. 1) show that color changes quite sharply as a function of both these parameters. Thus a central challenge has emerged for the halo mass quenching picture, namely, why do galaxy structural parameters predict the outcome of halo mass quenching better than halo mass itself does?  We return to this question below.

While correlations do not necessarily imply causality, they are strong hints, all of which has led to a rather complicated picture of galaxy evolution. Quenching may well involve a mix of complex processes that are likely to be dependent on several parameters that are themselves correlated. However, several themes emerge from the results discussed. The conditions of the bulge and perhaps the very center of the galaxy appear to be important. Indeed, \cite{kauffmann06} suggested that bulge-building is the underlying cause of their correlation between color and surface mass density. And several authors, cited above, concluded that high $n$ is necessary to quench a galaxy, providing further evidence for bulges. Moreover, since a hallmark of bulges is high central density, it is notable that \cite{wake12} find that central velocity dispersion is the single most correlated parameter of all with galaxy color. And finally, since bulges and high central densities are closely associated with black holes \citep{magorrian98, gebhardt00}, it is tempting to conclude that this mounting chain of evidence is simply a ``smoking gun'' pointing to AGN feedback. 

In total, these works suggest that internal processes, and specifically {\it central} processes, are responsible for shutting down star formation. As noted, this poses a problem at first sight for halo mass quenching, since halo properties are seen to correlate more weakly with quenching than do variables such as S\'ersic index and central velocity dispersion. However, a key element in the halo picture is radio mode, which depends on AGN feedback \citep{dekel06, croton06}, and thus possibly on internal/central conditions. Perhaps it will be possible to link these various processes in a plausible causative chain that explains all of the data. We return to this possibility in the discussion section. 

In this paper, we build on previous works and consider the possibility of multiple physical processes acting together in concert to quench star formation in galaxies.  Whereas most works have only explored global structural parameters, we explore both global and {\it central} structural parameters. Our data set is AEGIS galaxies possessing $HST$/ACS imaging, similar to \cite{mendez11} but over a narrower redshift range, $0.5 \le z < 0.8$. Like them, we use color as a proxy for quenching and structural parameters derived from the same GIM2D fits. However, we focus on different structural parameters and, importantly, convert luminous quantities of subcomponents to stellar mass using color-derived $M/L$ ratios. 

Our ultimate goal is to identify that parameter, or combination of parameters, that seems to be the best discriminant between star-forming and quenched galaxies.  Having found that combination, we compare its efficacy (or sharpness) to studies focusing on halo parameters \citep[e.g.,][]{woo12} in order to assess whether the primary driver of quenching is conditions that exist inside a galaxy or outside it.

A major result of this paper is that the S\'ersic index, $n$, displays the sharpest break between star-forming and quenched galaxies, i.e., it looks most like a quenching threshold. However, $n$ does not really distinguish red and blue galaxies all that well -- $\approx40\%$ of AEGIS galaxies with high $n$ have blue colors. Suspecting contamination from starbursts, AGN, or errors of measurement, we introduce a novel parameter that is closely related to $n$ but is more robustly measured, namely, central surface mass density, $\Sigma_{\rm 1 kpc}^*$. Under this parameter, we find that the number of outliers is dramatically reduced, implying that the innermost
structure of galaxies may be most fundamentally related to quenching.  Moreover, using stellar mass measurements of the bulge and inner 1 kpc region of galaxies, we
show that at $z\sim0.65$, most blue galaxies cannot simply fade onto the red sequence; they must instead undergo a {\it significant restructuring} of their innermost stellar density profiles en route to quenching. 

These results are compared to various theoretical models.  The first major conclusion is that the quenching sharpness found with our new parameter, central surface mass density, far exceeds that found with halo mass, highlighting a major tension with the halo quenching picture.  Looking at alternative theories, we find several striking 
points of agreement with the major-merger picture, but also some important caveats. These concerns suggest that bulge-building may just proceed quite naturally because galaxies at these redshifts are not yet very axisymmetric and non-central torques are constantly being generated. Finally, the very close connection between quenching state 
and central conditions that we find in this paper looks like a ``smoking gun'' for AGN feedback, yet not all aspects of the data are fully explained by that model either.
  
Finally, we place online\footnote{{\url{http://people.ucsc.edu/~echeung1/data.html}}} one of the most comprehensive datasets available, comprised of 11,223 galaxies at $0.2 < z < 1.2$, with a mean redshift of $z\sim.75$. One powerful aspect of this dataset is the use of multi-color $HST$/ACS $V$- and $I$-band imaging, which allows the accurate conversion of light to stellar mass. It also includes GIM2D bulge-disk decompositions \citep{simard02}, which provide photometric and structural measurements of the bulges and disks separately; these intermediate redshift galaxy decompositions are only possible thanks to the high resolution $HST$ imaging. Additionally, stellar masses are derived for the subcomponents using their $V,I$ colors.

This paper is organized as follows: \S 2 describes our data and derivation of the analyzed quantities. In \S 3, we explain our sample selection criteria and discuss sample completeness. \S 4 presents our main results -- the correlations between structural parameters and color. In \S 5, we compare our results with several theoretical models and present our two-stage scenario of galaxy evolution. Finally, we list our conclusions in \S 6. A cosmology with $H_{0} = 70$ km s$^{-1}$ Mpc$^{-1}$, $\Omega_{m} = 0.30$ and  $\Omega_{\Lambda} =0.70$ is used throughout this paper. All magnitudes are on the AB system. 

\section{Data} \label{sec:data}

We start with a description of all the main sources of data used in this paper, which come from AEGIS, and then discuss the sample selection in \S \ref{sec:selection}. For an overview of the AEGIS data, please see \cite{davis07}.

\subsection{CFHT $BRI$ Photometric Catalog} \label{sub:cfht}

The fist photometric data are we use from the Canada-France-Hawaii Telescope (CFHT) $BRI$ 
imaging catalog. The CFHT 12k camera has a 12,288 $\times$ 8192 pixel 
CCD mosaic array and a plate scale of 0.207\arcsec ~per pixel,
providing a field of view of $0.70^\circ \times 0.47^\circ$. 
Five separate fields, with one to five distinct CFHT 12k
pointings per field, were observed from 1999 September to 2000 October. 
The integration time for each points was $\sim1$ hour in $B$ and $R$ and for 
$\sim2$ hours in $I$, broken down to individual exposures of 600 s. The data are complete to $\sim25.25$ in $B$, $24.75$ in $R$, and $\sim24.25$ in $I$ (\citealp[see][]{coil04} for more details).

These $BRI$ magnitudes were used with $k$-correct v4.2 \citep{blantonroweis07} to obtain the rest-frame color ($U-B$) and absolute magnitudes ($M_B$) used throughout this paper.

\subsection{HST ACS $V$+$I$ Imaging and SExtractor Photometry}  \label{sub:hst/acs}
The main photometric catalog from which the sample was 
selected is based on $HST$/ACS images taken as part of the
AEGIS survey \citep{davis07} under program GO-10134 (PI: M. Davis).  The
exposures were taken between 2004 June and 2005 March over 63
tiles covering an area approximately 10.1\arcmin $\times$ 70.5\arcmin~
in size. Each tile was observed for a single orbit in 
F606W ($V$) and F814W ($I$) using a four-point dither pattern. 
These pointings were combined with the STSDAS Multidrizzle
package using a square kernel. The final images have a pixel scale of
0.03\arcsec\ per pixel and a point-spread function (PSF) of 0.12\arcsec\
FWHM. The 5$\sigma$ limiting magnitudes for a point source are
$V=28.14$ and $I=27.52$ within a circular aperture of radius
0.12\arcsec($\sim50$-pixel area). For an extended object, the
5$\sigma$ limiting magnitudes are $V=26.23$ and $I=25.61$
for a circular aperture of radius 0.3\arcsec ($\sim314$ pixel area).

SExtractor \citep{bertin96} is used to detect objects in summed ACS $V$+$I$ images and to construct initial galaxy segmentation maps. A detection threshold of 1.5$\sigma$ and
50 pixels is chosen. These detection maps and the ACS zero points
\citep{sirianni05} were applied to each band separately to create the
ACS photometric catalogs. We selected all nonstellar objects with
SExtractor CLASS\_STAR $< 0.9$ and $I<25.0$ that did not lie within 50
pixels of a tile edge for our automated morphology analysis, covering
an effective area of 710.9 $\rm{arcmin}^2$ in the ACS
images \citep[see][for more details]{lotz08}.

This high resolution catalog was used to generate the galaxy sample comprising the GIM2D bulge+disk catalogs. 

\subsection{GIM2D} \label{sub:gim2d}

Structural parameters of the $HST$/ACS imaged galaxies were measured 
using GIM2D, a 2D bulge+disk decomposition
program \citep{simard02}. Three separate fits were made:
a single S\'ersic fit with floating $n$, a bulge + disk
fit with $n_{bulge} = 2$ and $n_{disk} = 1$, and a
bulge + disk fit with $n_{bulge} = 4$ and $n_{disk} = 1$
The three fits were done
simultaneously using both the $V$ and $I$ $HST$/ACS
images according to the procedure in \cite{simard02}.
The bulge surface brightness profile is
parameterized by:
 
\begin{equation}
\Sigma(r) = \Sigma_{\rm e}\exp\{k[(r/r_{\rm e})^{1/n} -1]\}, \label{eqn:sersic}
\end{equation}

\noindent
as given by \citep{sersic68}. Here, the 
parameter $k$ is set equal to $1.9992n-0.3271$,
so $r_{\rm e}$ remains the projected radius enclosing half the
light \citep{capaccioli89}. The disk profile is a simple exponential:

\begin{equation}
\Sigma(r) = \Sigma_{\rm 0}\exp(-r/r_{\rm d}), \label{eqn:sersicdisk}
\end{equation}

\noindent
where $\Sigma_{\rm 0}$ is the face-on central surface brightness and $r_{\rm d}$ is the semimajor axis scale length. 
For the single S\'ersic fit, Eqn.~\ref{eqn:sersic} is used to fit the whole galaxy. 

GIM2D also measures concentration, which, unlike the SDSS definition, is defined as the ratio of the inner and outer isophote fluxes of normalized radii $\alpha$ and 1; we follow \cite{abraham94} and use $\alpha=0.3$. Additionally, the GIM2D models produce galaxy, bulge, and disk $V,I$ magnitudes -- these are the primary magnitudes used throughout this paper.

Throughout this paper, most of our results
utilize the single S\'ersic fits (Table~\ref{tab:3}).
When examining the bulge and disk properties,
we use the best-fitting, two-component
decomposition, i.e., either $n_{bulge}=4$ or $n_{bulge}=2$ (Tables~\ref{tab:4} and \ref{tab:5}), for
each galaxy as indicated by $\chi^{2}$. We only use bulge measurements of $B/T>0.1$ galaxies, as the low signal-to-noise makes measurements of systems with $B/T<0.1$ uncertain. Our subcomponent sample with GIM2D measurements of the bulge and disk separately
is comprised of $\approx 60\%$ from the $n_{bulge}=2$ and $\approx 40\%$ from 
the $n_{bulge}=4$ fits. Comparing the $n_{bulge}=4$ fit to the $n_{bulge}=2$ fit shows a median offset of $\log~M_{\rm *,bulge}$ to be $\approx 0.10$ dex with
a dispersion of $\approx 0.23$ dex while $\log~r_{\rm e,bulge}$ has a median offset of $\approx 0.15$ dex with a dispersion of $\approx 0.28$ dex; both parameters are offset toward higher values in the $n_{bulge}=4$ fit.

\subsection{DEEP2 + DEEP3 Redshift Survey} \label{sub:specz}

Spectroscopic redshifts were measured in the DEEP2 redshift
survey using the DEIMOS spectrograph \citep{faber03}
on the Keck II telescope \citep{davis03, newman12}. Targets were selected for DEEP2 spectroscopy from the CFHT $BRI$ imaging described in \S\ref{sub:cfht}. Most of DEEP2 used the $BRI$ photometry to screen out low-redshift galaxies, but this screening
was not applied in the AEGIS region, and so the resulting sample
is representative from $z = 0$ to $z \sim 1.4$. Eligible targets must have $18.5\le R \le 24.1$ and surface brightness brighter than $\mu_{R} \le 26.5$ \citep{davis03, newman12}.

Additional spectroscopic redshifts are available in the recently completed DEEP3 redshift survey \citep{cooper11, cooper12}. This survey shares many of the same characteristics of DEEP2, i.e., they both use the DEIMOS spectrograph and were both preselected using CFHT $BRI$ photometry. However, while DEEP2 used a 1200 line/mm grating in DEIMOS, DEEP3 employed a 600 line/mm grating, resulting in spectra of lower resolution. The quality of the redshifts, however, are unaffected.  

Taking only spectroscopic redshifts with quality code of $Q=3$ or $Q=4$ and cross matching it to the $HST$/ACS catalog yields a sample of 6310 galaxies; these galaxies make up the spectroscopic sample.

\subsection{Photometric Redshifts} \label{sub:photoz}

The DEEP2+DEEP3 survey is approximately 65\% complete to $R = 24.1$ in AEGIS \citep{newman12}. For those galaxies without spectroscopic $z$ and to extend the sample to fainter limits, we utilized photometric redshifts (J. Huang et al., in prep) derived from the Artificial Neural Networks method (ANNz, \citealt{collister04}) using the multi-wavelength AEGIS photometry that includes 12 unique bands in the wavelength range from $u$ to 8 $\mu$m, with deep {\it Spitzer}/IRAC photometry (\citealt{davis07, barmby08}; Zheng et al., in prep) as the base. This sample was 3.6 $\mu$m selected ($f_{3.6} > 2 \mu$Jy) with a color cut to isolate $z<1.5$ galaxies. The redshift catalog is complete down to $\log~M_*/M_{\odot}=9.5$ for $0.4< z < 1.2$, and the rms accuracy is $\Delta z/(1+z) = 0.025$. Cross matching this sample to the $HST$/ACS catalog that does not have a quality spectroscopic redshift yields 4913 galaxies; these galaxies make up the photometric sample. The total number of galaxies in our initial sample, consisting of the spectroscopic and photometric sample, is 11,223

\subsection{Rest-frame Absolute $B$ Magnitudes and $U-B$ Colors} \label{sub:rest-frame}

Rest-frame absolute $M_B$ magnitudes and $U-B$ colors are needed for both integrated galaxies and for bulge and disk subcomponents separately. For galaxies, these quantities are obtained through $k$-correct v4.2 ($k$-corrected down to $z=0$; \citealp{blantonroweis07}) with CFHT $BRI$ photometry and redshift as inputs. 

For bulges and disks, however, CFHT $BRI$ photometry is not available, but there is $HST$/ACS $V$ and $I$ photometry modeled by GIM2D. In order to be consistent with the galaxy values, we derive a calibration for $M_B$ and $U-B$ from $V$, $I$, and redshift. We use the galaxy rest-frame magnitudes from $k$-correct as fiducial values to derive this calibration, which was then used to calculate $M_B$ and $U-B$ for the subcomponents sample.

The functional form we use for $M_B$ is \citep{gebhardt03}\footnote{We use \cite{gebhardt03}'s form but fit for our own coefficients}:

\begin{eqnarray}
M_{B} = I_{814} - DM(\Omega_{m}, \Omega_{\Lambda}, \Omega_{K}) + K_{IB}, \label{eqn:mb}
\end{eqnarray}

\noindent
where DM is the distance modulus for the adopted cosmology and

\begin{eqnarray}
K_{IB} &=& 1.490 - 18.266z + 94.056z^{2}  \nonumber \\
&& {} - 229.782z^{3} + 294.741z^{4} - 189.892z^{5} + 48.034z^{6} \nonumber \\
&& {} + (2.233 - 5.448z + 3.187z^{2} - 0.082z^{3})(V-I) \nonumber \\
&& {}+ (0.592 - 0.540z - 0.036z^{2})(V-I)^{2}. \label{eqn:kcorrect}
\end{eqnarray} 

The functional form for $U-B$ is \citep{gebhardt03}:

\begin{eqnarray}
U-B &=&-0.882 + 16.627z - 84.798z^{2} \nonumber \\ 
&& {} + 212.831z^{3} - 286.211z^{4} + 195.256z^{5} - 52.584z^{6} \nonumber \\
&& {} + (0.492 + 0.380z + 0.415z^{2} - 0.493z^{3})(V-I) \nonumber \\
&&+ (0.751 - 1.609z + 0.739z^{2})(V-I)^{2} \label{eqn:ub}
\end{eqnarray} 

Fig.~\ref{fig:compare_all}a and \ref{fig:compare_all}b compares the values of $M_B$
and $U-B$ from Eqn. \ref{eqn:mb}-\ref{eqn:ub} to those derived from $k$-correct. The relations are nicely linear with $\sigma(M_B) = 0.215$ mag and $\sigma(U-B) = 0.096$ mag. We use these relations to compute $M_B$ and $U-B$ for the subcomponent sample. We also use these equations for the galaxies in our sample that have ill-measured CFHT $BRI$ measurements, characterized by large errors ($\approx 7\%$).

\subsection{Stellar Masses} \label{sub:mass}

Stellar masses for most of our sample are available from J. Huang et al., (in prep.). Using a Salpeter initial mass function (IMF), the multi-wavelength AEGIS photometry (with deep {\it Spitzer}/IRAC photometry as the base (\citealt{davis07,barmby08}; Zheng et al., in prep; J. Huang et al., in prep) were fit to a grid of synthetic SEDs from \cite{bruzualcharlot03}, assuming solar metallicity. These synthetic SEDs span a range of ages, dust content, and exponentially declining star formation histories. The typical widths of the stellar mass probability distributions are $0.1-0.2$ dex.

To obtain stellar masses for the subcomponent samples we utilize the well-known correlation between mass-to-light ratio $(M_*/L)$ and optical colors \citep[e.g.,][]{bell01}. To account for our large redshift range, we add a redshift-dependent term to the relationship, similar to the approach in \cite{lin07} and \cite{weiner09}. Options are to use either rest-frame $U-B$ (from $k$-correct) or observed $V-I$. To aid our choice, we make fits using both colors and compare them to the $M_*/L_B$ values in Fig.~\ref{fig:compare_ml}. The left panels display $M_*/L_B$ vs. observed $V -I$ while the right panels plot $M_*/L_B$ vs. rest-frame $U -B$. Each row represents a different redshift range. Overplotted in each panel is a red dashed curve that represents our fit. The fit for $M_*/L_B$ as a function of $V -I$ is a better match than the fit to $U-B$, especially at high redshifts. The final adopted expression for $M_*/L_B$ is:

\begin{eqnarray} 
\log ~M_*/L_{B} &=& -0.340 - 2.593z + 1.195z^{2}  \nonumber \\
&&  {} 1.908(V-I) - 0.432(V-I)^{2}  \label{eqn:masslight}
\end{eqnarray}

Together with the absolute $B$ magnitudes measured from Eqns.~\ref{eqn:mb} and \ref{eqn:kcorrect}, we are now in a position to calculate stellar masses for any object with a measured $V,I,$ and redshift. These calibrated fits are used to obtain $M_*$ values for the subcomponent samples. For the galaxies in our sample that do not have stellar masses from J. Huang et al., (in prep) ($\approx10\%$), their $M_*$ are also obtained this way; these stellar masses are shown in Table~\ref{tab:1}. We make a final check of our method by comparing these derived stellar masses to those calculated by J. Huang et al., (in prep); this is shown in Fig.~\ref{fig:compare_all}c, where the relationship is well-behaved with an rms scatter of $0.175$ dex.

\subsection{Error Estimates} \label{sub:errors}

\begin{figure}[t!] 
\centering
\includegraphics[width=.48\textwidth]{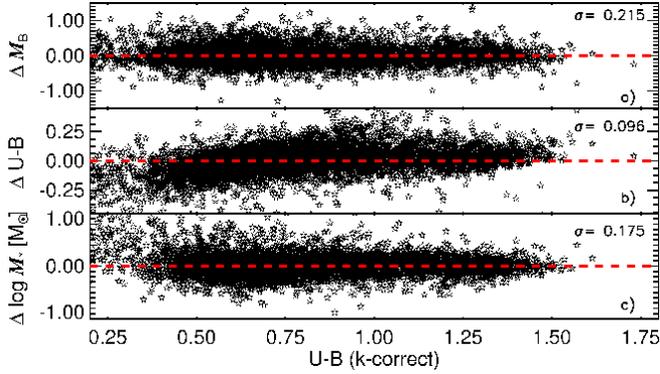}
\caption{We plot the residuals of rest-frame absolute $B$-band magnitude $M_{B}$, rest-frame $U-B$, and stellar mass $M_*$ against rest-frame $U-B$ from $k$-correct. Panel a shows the difference between $M_{B}$ computed from $k$-correct v4.2 \citep{blantonroweis07} using CFHT $BRI$ and redshift and $M_{B}$ computed from Eqn.~\ref{eqn:mb} and~\ref{eqn:kcorrect}. Panel b plots the residuals of $U-B$ from $k$-correct to $U-B$ from Eqn.~\ref{eqn:ub}. Panel c shows the residuals of $M_*$ from J. Huang et al., (in prep) and $M_*$ obtained using Eqn.~\ref{eqn:mb} and~ \ref{eqn:masslight}. From the one-to-one red dashed line in each panel, it is clear that values derived from our fits are consistent with the fiducial values. The dispersion $\sigma$ is displayed on the top of each panel.
\label{fig:compare_all}}
\end{figure}

\begin{figure}[t!] 
\centering
\includegraphics[width=.45\textwidth]{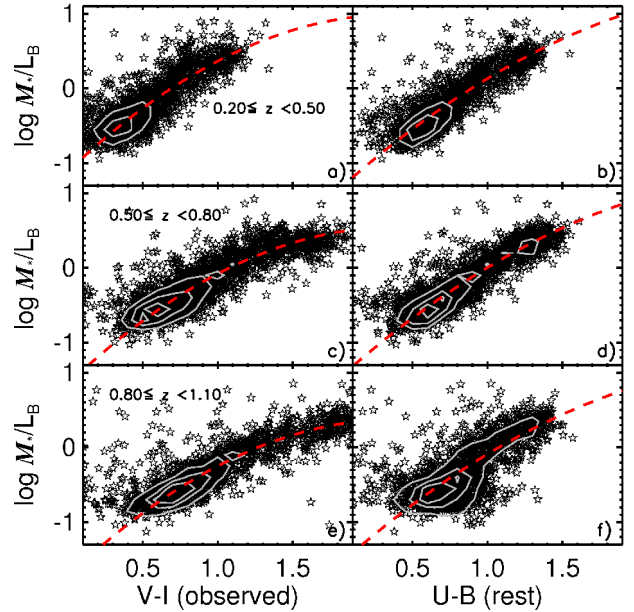}
\caption{$M_*/L_B$ ratio (mass from J. Huang et al., in prep and $B$-band luminosity from $k$-correct) vs. observed $V-I$ (left column) and rest frame $U-B$ (right column) in three redshift bins: a)-b) $0.20 \le z < 0.50$, c)-d) $0.50 \le z<0.80$, and e)-f) $0.80 \le z<1.10$. Contours are shown to give a sense of the relative number densities. The red dashed curve in each panel represents our fit; the fit for $M_*/L_B$ as a function of observed $V-I$ is better than the fit to rest-frame $U-B$, hence we adopt it.
\label{fig:compare_ml}}
\end{figure}

All error estimates measured by GIM2D, i.e., the structural parameters such as $r_{\rm e}$ and $n$, are $99\%$ confidence limits \citep{simard02}; we convert these into 1-$\sigma$ limits assuming a Gaussian distribution. There are two sources of stellar mass: those from the SED-fitting and those from our mass fits. The errors for the former are the typical widths of the stellar mass probability distribution ($0.1-0.2$ dex). The errors for masses obtained from Eqn.~\ref{eqn:masslight},~\ref{eqn:mb}, and~\ref{eqn:kcorrect} are the standard deviation of the residual distribution between the fitted masses and those of J. Huang et al., (in prep) (see Fig.~\ref{fig:compare_all}c; $\sigma\approx 0.175$).  Errors for $U-B$ and $M_B$ obtained from $k$-correct are estimated by measuring the 1-$\sigma$ dispersion from the $HST$/ACS $V$ and $I$ in the redshift range $0.64 < z <0.68$ and $0.82<z<0.86$, respectively. Within these redshift ranges, rest-frame $U$ and $B$ approximately redshifts into observed $V$ and $I$, which when combined with the high resolution of $HST$, gives us an accurate photometric error estimate. The average errors are $\approx 0.07$ mag ($U$) and $\approx 0.05$ mag ($B$). Errors for $U-B$ and $M_B$ obtained from Eqn.~\ref{eqn:mb} and \ref{eqn:ub} are taken to be the standard deviation between our fits and $k$-correct. For the mass-radius combinations, e.g., $M_*/r_{\rm e}$, we propagate the errors from the masses and the GIM2D confidence limits. 

\begin{figure}[t!] 
\centering
\includegraphics[width=.35\textwidth,angle=90]{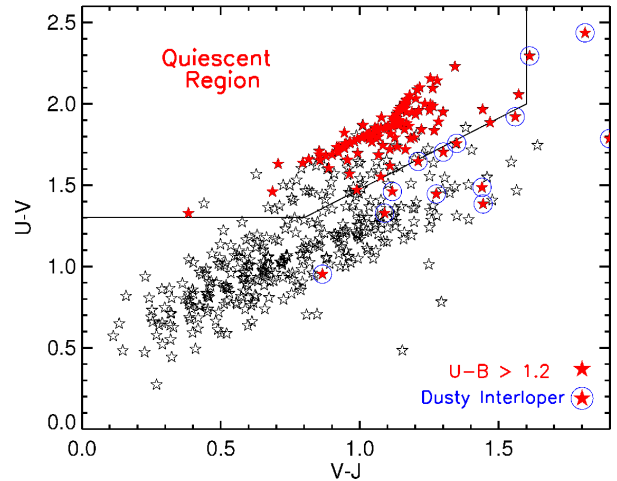}
\caption{To address dust, we plot our sample in the rest-frame $UVJ$ diagram. The quiescent population lie within the quiescent region as defined by \cite{williams09}. Red points represent red sequence galaxies, defined to have $U-B>1.20$. Almost all of the red sequence lies within the quiescent area. $U-B>1.20$ galaxies outside the quiescent area is only $\approx3\%$ of the defined red sequence. These are the dusty star-formers, and we eliminate them from the sample so that the red sequence galaxies are truly quiescent. 
\label{fig:uvj}}
\end{figure}

\begin{figure*}[t!] 
\centering
\includegraphics[scale=.60,angle=90]{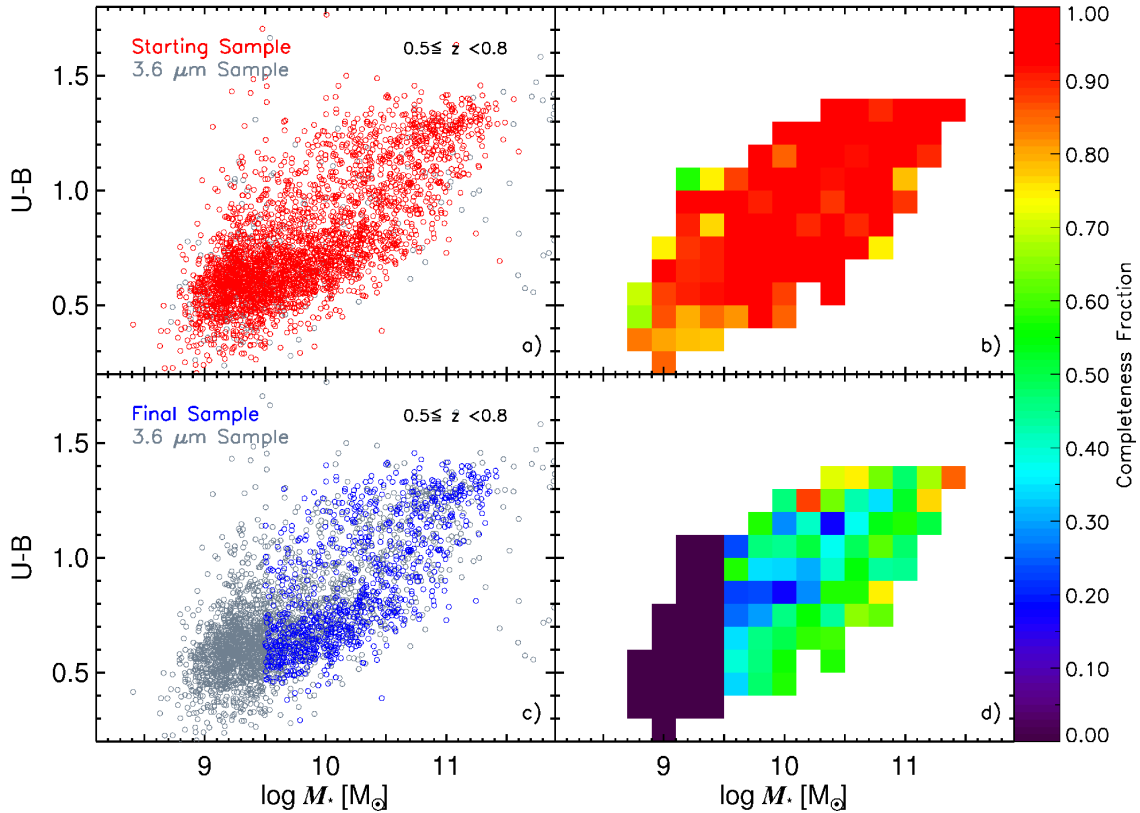}
\caption{$Left~column:$ Rest-frame $U-B$ vs. $\log~M_*$ is plotted for the `starting' sample (red; top row) and the `final' sample (blue; bottom row). For comparison, the $Spitzer$/IRAC 3.6 $\mu$m-selected sample (J. Huang et al., in prep) in our redshift range $0.5\le z < 0.8$ is plotted in the background. $Right~column:$ $U-B$ vs. $\log~M_*$ is binned with lengths in $U-B$ and $\log~M_*$ that correspond roughly to their  1$-\sigma$ error; we only show bins with more than 5 galaxies. Within each bin, the fraction of the number of galaxies in that row's sample to that of the 3.6 $\mu$m-selected sample is computed and displayed as the corresponding color indicated by the color bar to the right. The completeness of the `starting' sample is uniformly complete above $\log~M_*/M_{\odot}= 9.5$. The completeness of the `final' sample is $\sim50\%$ and is largely uniform. See text for discussion.
\label{fig:selection_effects}}
\end{figure*}

\section{Sample Selection} \label{sec:selection}

Within the AEGIS region, $\approx 30,000$ objects have both $HST/$ACS imaging and GIM2D decompositions; this is the master GIM2D sample. Only 11,223, however, have either a spectroscopic or photometric redshift (see \S\ref{sec:data}). Moreover, although our redshift coverage is from $0.2 < z < 1.2$, in order to minimize $k$-corrections, we restrict our sample to $0.5 \le z < 0.8$; this cuts our sample down to 3,426, this will be referred to as the `starting' sample. To reduce the effects of dust, we only choose galaxies with axis ratios ($b/a$; as measured from the single $n$ fit) greater than 0.55, furthering reducing our sample to 1,567 galaxies. 
 
Although GIM2D was run for every galaxy, not every decomposition is reliable. For example, galaxies with effective radii $r_{\rm e}$ less than half the full width at half-maximum (FWHM) of the point-spread function (PSF; 2 pixels) are not well fit. Additionally, galaxy models created by GIM2D that are offset from the center of the $HST$/ACS image by more than 3.5 pixels are similarly ill-fit. There are also instances where the fitting failed; eliminating these leaves us with 1,427 objects. Note that we are only using the single S\'ersic fit values for global galaxy parameters, and hence this sample consists of values only from the single $n$ fit.  
  
GIM2D was also used to produce measurements of every galaxy's bulge and disk through two different fits -- the $n_{bulge}=4$ and $n_{bulge}=2$ decompositions with the disk being $n=1$ for both (see \S \ref{sub:gim2d}). Note that GIM2D bulge+disk decompositions do allow for a galaxy to have $B/T=0$, i.e., a pure disk galaxy, if that is the optimal fit according to the Metropolis fitting algorithm ($\approx22\%$ of the subcomponent sample have $B/T=0$; see \citealp{simard02} for more details). For each galaxy, we use the bulge+disk fit with the smallest $\chi^2$. We only use the subcomponent measurements of the `final' sample, which we define below. 

To reduce the effects of dust, we applied an axis ratio cut of $b/a > 0.55$.. While this cut eliminates many edge-on dusty galaxies \citep{martin07}, it does not affect dusty face-on galaxies. To clean these from out sample, we calculate $UVJ$ rest-frame magnitudes. The resultant $U-V$ vs. $V-J$ two color plot enable us to separate dusty red galaxies from truly quiescent red galaxies \citep{williams09}. We use the $Rainbow$ software described in \cite{barro11a, barro11b}. Briefly, the software applies a $\chi^2$ minimization algorithm to find the best fitting galaxy template from the multi-wavelength photometry of AEGIS. Then several filters ($U$ Bessel, $V$ Bessel, $J$ Johnson) are convolved with the best template to estimate synthetic fluxes assuming a luminosity distance of 10 pc. Our results can be seen in the $UVJ$ diagram (Fig.~\ref{fig:uvj}). The upper-left region bounded by the solid lines within the $UVJ$ diagram represents the quiescent region, as defined by \cite{williams09}. Comparing the quiescent galaxies to the red sequence galaxies, which we define to be galaxies with $U-B>1.20$ and are shown in red in Fig.~\ref{fig:uvj}, shows excellent agreement; only 17 \footnote{4 of these galaxies are not visible because they have $UVJ$ magnitudes that are identical to those that are visible.} ($6\%$) of the $U-B>1.20$ galaxies lie outside the quiescent area. These are presumed to be dusty, star-formers and are discarded from our sample. There is an additional reduction of 8 $U-B>1.20$ galaxies because we were unable to obtain their $UVJ$ magnitudes. Since we do not know whether these galaxies are truly quiescent or simply dusty, we take the conservative route and discard them.  To sum, we require our quiescent galaxies to have $U-B>1.20$ {\it and} to lie within the quiescent region of the $UVJ$ diagram. With this criterion, our galaxy sample has 1,402 galaxies. 

\subsection{Completeness} \label{sub:complete}

Finally, we must discuss our sample's completeness. Because the DEEP2+3 spectroscopic survey is limited by an $R$-band magnitude of 24.1\footnote{There are some DEEP3 targets fainter than this limit \citep{cooper11,cooper12} }, there is a selection bias against low-mass galaxies. Fortunately, the photometric sample goes deeper, down to an IRAC $3.6~\mu$m flux of $2~\mu$Jy. Details of the photometric sample can be found in J. Huang et al., (in prep), but we will briefly summarize the key characteristics. The $3.6~\mu$m-selected sample spans the redshift range of $0.4 < z < 1.2$, where $3.6~\mu$m also probes the rest-frame NIR ($1.2-2.5~\mu$m). Galaxies of all types have very similar SEDs in the NIR band. Therefore a rest-frame NIR-selected sample suffers no bias against either blue or red galaxies \citep{cowie96, huang97}. Galaxy NIR luminosities also trace their underlying stellar mass, in other words, this sample is very close to a mass-selected sample. The $K$-band absolute magnitudes for galaxies in this sample are calculated with the $3.6~\mu$m flux densities. The IRAC-to-$K$-band $k$-correction is adopted from \cite{depropris07}. The absolute $K$-band magnitude range for this sample is $-19 < M_K < -25$. This translates into a limiting stellar mass of $\log~M_*/M_\odot=9.5$. Therefore, cross-matching to the photometric sample has essentially eliminated the selection bias against low-mass galaxies of the DEEP2+3 surveys. 

To illustrate our sample completeness, we compare the color-mass diagrams of our `starting' sample (red) to the $Spitzer$/IRAC $3.6~\mu$m-selected sample (gray) in Fig.~\ref{fig:selection_effects}a. There are hardly any gray points, indicating that the `starting' sample contains almost all the galaxies in the $3.6~\mu$-selected sample. This is further illustrated in Fig.~\ref{fig:selection_effects}b where we bin up the color-mass diagram with lengths in $U-B$ and $\log~M_*$ that roughly correspond to their distributions'  1$-\sigma$ error; we only show bins with more than 5 galaxies. Within each bin, the fraction of the number of galaxies in the `starting' sample to that of the 3.6 $\mu$m-selected sample is computed and displayed as the corresponding color indicated by the color bar to the right. Confirming what was seen in Fig.~\ref{fig:selection_effects}a, the completeness is almost perfect, and most importantly, the completeness is uniform, especially for $\log~M_*/M_\odot>9.5$, the mass limit of the 3.6 $\mu$m-selected sample. Thus, our `starting' sample is uniformly complete down to the mass limit of the 3.6 $\mu$m-selected sample. 

However, our `starting' sample is not the ultimate sample we use. To get rid of bad data and dusty galaxies, we have imposed several requirements (see \S\ref{sec:selection}). To obtain our `final' sample, we impose one final requirement, $\log~M_*/M_\odot>9.5$. This last cut ensures that our `final' sample is complete above $\log~M_*/M_{\odot}=9.5$. Thus finally, we have our `final' sample, consisting of 943 galaxies. The `final' sample is what is plotted in all subsequent figures unless stated otherwise. The completeness of the `final' sample is illustrated in Fig.~\ref{fig:selection_effects}c and Fig.~\ref{fig:selection_effects}d. Fig~\ref{fig:selection_effects}d (calculates bins of completeness like in Fig.~\ref{fig:selection_effects}b) shows that the completeness of the `final' sample is $\sim50\%$, with a dearth of galaxies on the top of the blue cloud, i.e., the green valley, and a surplus of galaxies on the upper red sequence. These features are due to the $b/a$ criterion, which is meant to eliminate edge-on galaxies that are presumably dusty. Indeed, according to \cite{martin07}, dusty galaxies do primarily reside on top of the blue cloud, which explains why there is a lack of galaxies on top of the blue cloud in the `final' sample compared to the 3.6 $\mu$m sample. The surplus of galaxies on top of the red sequence is also understandable since the reddest galaxies are elliptical galaxies that have intrinsically high axis ratios. Although there are some biases introduced into the `final' sample by these various cuts, we have tested the effects of removing them and find that it does not affect our conclusions. But we stress that these cuts are {\it necessary}; they remove bad data. Our `final' sample is a culmination of the best data from our available resources.  For an extra discussion of our samples' surface brightness limits, data quality, and possible S\'ersic index bias, please see Appendix \ref{app:sb}, \ref{app:gim2derror}, \ref{app:sersicindex}.

All our data, including those that were not presented in this paper, are available online at: \url{http://people.ucsc.edu/~echeung1/data.html}. Tables~\ref{tab:1}-\ref{tab:5} present the key parameters we use in our paper for twenty randomly selected galaxies in our catalog. Table \ref{tab:1} presents basic information of our galaxies, including their unique IDs, derived photometric quantities, and stellar masses both from $k$-correct and Eqn.~\ref{eqn:masslight}, \ref{eqn:mb}, and \ref{eqn:kcorrect}. Table \ref{tab:2} presents much of the same information in Table 1, but only for the subcomponents. Tables \ref{tab:3},~\ref{tab:4}, and~\ref{tab:5} present the three GIM2D catalogs: the single S\'ersic fit, $n_{bulge}=4$ fit, and $n_{bulge}=2$ fit, respectively. These GIM2D catalogs provides many measurements, including $V$, $I$, and $r_{\rm e}$ for both the galaxy and its subcomponents. The rest of the measurements and galaxies can be obtained online. 

\section{Results}

\subsection{The Most Discriminating Color Parameter} \label{sub:best}

\begin{figure*}[t!]
 \centering
\includegraphics[scale=.70,angle=90]{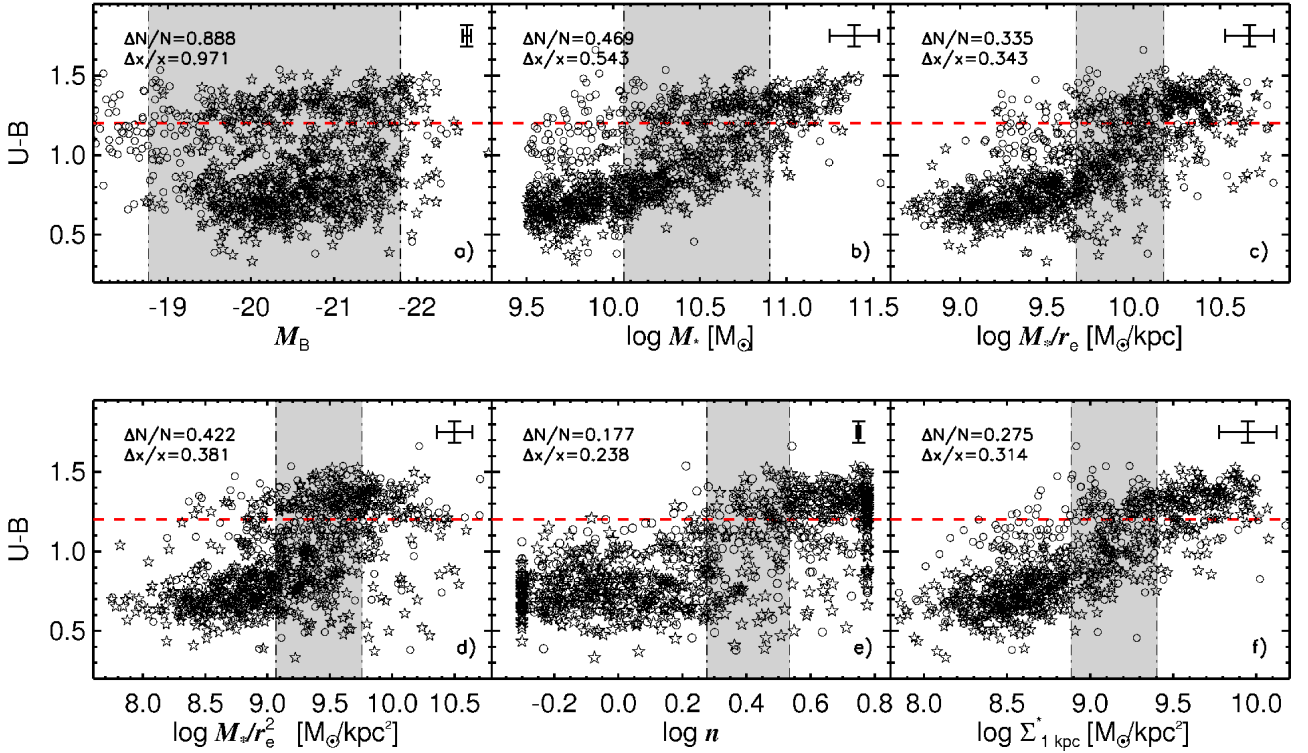}
\caption{Rest-frame $U-B$ color is plotted against: a) rest-frame absolute $B$-band magnitude $M_{B}$, b) stellar mass $M_*$, c) stellar mass divided by semimajor axis effective radius $M_{*}/r_{\rm e}$, d) $M_{*}/r_{\rm e}^2$ (surface density), e) S\'ersic index $n$, and f) stellar mass surface density within 1 kpc $\Sigma_{\rm 1 kpc}^*$.  $M_*$ is in solar masses and $r_{\rm e}$ is in kpc.  The stars and circles represent galaxies with spectroscopic redshifts and photometric redshifts, respectively. 1-$\sigma$ error bars are displayed in the upper right of each panel. The overlap region of each parameter is highlighted in gray.  Overlap regions are defined as regions in which the fraction of red galaxies is between $15\%-60\%$ (see \S\ref{sub:best}). The red horizontal line represents the division between red and blue galaxies.  The top left corner of each panel shows two measures of the size of the overlap region: the fractional number of galaxies within the overlap region $\Delta$N/N and the fractional extent of the region $\Delta$x/x, where x contains 90\% of the points (see text). This analysis was done on the `final' sample, which is complete only in stellar mass (down to $\log~M_*/M_{\odot}=9.5$; see \S\ref{sub:complete}). By both measures, $M_*/r_{\rm e}$ outperforms other parameters of the form $M_*/r_{\rm e}^p$, which we have confirmed by studying intermediate values of $p$.  S\'ersic index and $\Sigma^*_{\rm 1 kpc}$ do even better, hinting that the distribution of mass in the inner parts of galaxies may play a fundamental role in quenching star formation.
 \label{fig:findcrit_all}}
 \end{figure*}

We begin by intercomparing the various {\it global} structural parameters discussed in the introduction to find which is the best predictor of color. Fig.~\ref{fig:findcrit_all} plots $U-B$ rest-frame color against six quantities for the final galaxy sample: rest-frame absolute $B$-band magnitude $M_{B}$, stellar mass $M_*$, stellar mass divided by semimajor axis effective radius $M_{*}/r_{\rm e}$ (sometimes called the ``inferred velocity dispersion'')\footnote{The true stellar velocity dispersion is $\sigma^2 \propto GM/r_{\rm e}$, where $M$ is the total mass including stars, gas, and dark matter. \cite{franx08} provide a value of the coefficient through the fitting of a sample of SDSS galaxies: $\sigma^2 = 0.3GM_*/r_{\rm e}$. Recently, \cite{taylor11} and \cite{bezanson11} showed that the addition of a S\'ersic dependent term to the ``inferred velocity dispersion'' of \cite{franx08} provides a better estimate of the true velocity dispersion. We choose not to use this updated ``inferred velocity dispersion'' because we want to compare the color correlations of these parameters independently.}, $M_{*}/r_{\rm e}^2$ (nominal surface density)\footnote{Surface mass density is actually $M_{*}/2\pi r_{\rm e}^2$, but we omit the constants.}, S\'ersic index $n$, and inner stellar mass surface density $\Sigma_{\rm 1 kpc}^*$ (we defer discussion of $\Sigma_{\rm 1 kpc}^*$ to \S\ref{sub:surfacedensity}). The 1-$\sigma$ error bars are given in the top right of each panel (see \S\ref{sub:errors} for details). The spectroscopic sample and photometric sample are shown in open stars and open circles, respectively; we use this scheme throughout the rest of the paper. As stated in the introduction, the amount of color overlap is one measure of how well a parameter separates red sequence and blue cloud galaxies, and parameters that reduce this overlap are better discriminators of galaxy quenching. The sample considered is the `final' sample, which is complete only in stellar mass, as defined in \S\ref{sec:selection}. Thus the results of the following analysis is only applicable for the `final' sample. The goal of this section is to quantify the amount of overlap in order to determine the single best color discriminant among the traditional parameters.

As in \cite{strateva01}, the color-magnitude diagram (Fig. 5a) shows a clear red sequence and blue cloud. However, these two groups overlap greatly over the entire range of absolute magnitude. This confirms the well-known result that the $B$-band magnitude is a poor predictor of galaxy color.

Fig.~\ref{fig:findcrit_all}b shows the color-mass diagram. As shown by \cite{kauffmann03} and \cite{borch06}, mass is better correlated with star-formation history than is luminosity. Although the relationship with color is improved, the range of color overlap is still large, extending over $\sim0.8$ dex in mass. Fig.~\ref{fig:findcrit_all}c and \ref{fig:findcrit_all}d add powers of $r_{\rm e}$ in the denominator to $M_*$, in the form of $M_*/r_{\rm e}^p$.  The smallest overlap by eye is given by $M_*/r_{\rm e}$ in Fig.~\ref{fig:findcrit_all}c, while mass surface density $M_*/r_{\rm e}^2$ in Fig.~\ref{fig:findcrit_all}d looks slightly worse. Thus, this new DEEP2 sample indicates that $M_*/r_{\rm e}$ is a superior color discriminator to surface mass density $M_*/r_{\rm e}^2$.

To summarize, effective radius $r_{\rm e}$ tightens the basic color-mass relation because red galaxies at fixed mass are smaller than blue galaxies, and the tightest correlation is obtained using $M_*/r_{\rm e}$. 

We plot in Fig.~\ref{fig:findcrit_all}e color vs. S\'ersic index $n$. The character of this plot is markedly different from the others -- rather than a smooth trend with color within the blue cloud as in, for example, $M_*/r_{\rm e}$, the color jump is more abrupt, with color remaining constant above and below what appears to be a critical value of $n$ around $\log~n =0.36$ ($n=2.3$). This behavior is intriguing because it might signal a {\it real physical threshold} in S\'ersic index, above which star formation shuts down. As stated in the introduction, $n$ likely plays an important role in quenching star-formation. \cite{blanton03} and \cite{schiminovich07} demonstrated a trend between $n$ and color for SDSS galaxies. \cite{driver06} and \cite{allen06} also showed this relationship with their Millennium catalog. And \cite{wuyts11} demonstrated that this relationship persists out to $z\sim2.5$. \cite{bell08} and \cite{bell12} explored this correlation and showed that high $n$ is necessary for quenching but not sufficient -- there are many galaxies that are blue despite having high $n$. We see something similar in our data with the scattering of aberrant points in the lower-right-hand corner of Fig.~\ref{fig:findcrit_all}e. We term these aberrant points ``outliers'' and discuss them further in \S\ref{sub:outliers}.  

The above conclusions are based mainly on visual assessment of Fig.~\ref{fig:findcrit_all}. To quantify our results, we now present two new quantities that are designed to measure the size of the overlap regions.  These measurements can be applied to rank the predictive power of the various structural parameters and also to identify galaxies within the overlap regions for further study.  To define these quantities, we first bin the sample by the parameter of interest. Then, within each bin, we find the fraction of galaxies that are red, i.e., galaxies with $U-B>1.20$, which we have ensured to be genuinely quiescent and not dusty (see \S\ref{sec:selection}). The locations where this fraction equals $15\%$ and $60\%$ mark the beginning and end of the overlap region, respectively; these percentages were adjusted to match the core of the overlap regions as judged by eye and are a compromise over all diagrams. We varied the overlap definition with various permutations of starting boundaries in the red fraction range of $5\%$ to $20\%$ and ending boundaries from $40\%$ to $60\%$ and found that the results are unchanged. To find the locations of the edges of the overlap regions, we fit a fourth-order polynomial to the red fraction bins and interpolate to find where the fit reaches the desired fractions.  Each parameter has been divided into 25 bins, and each edge value is examined to ensure that the choice is sensible.  The edge locations depend only weakly on the choice of bin width -- for example, in the case of $M_*/r_{\rm e}$, bin sizes in the range $0.07-0.25$ dex produce similar results.

We define two measures to quantify the sizes of the overlap regions. The first is the fractional number of galaxies in the region, $\Delta$N/N, where N is the total number of galaxies and $\Delta$N is the number within the overlap region.  The second is the fractional extent of the region, $\Delta$x/x, where $\Delta$x is the width of the overlap and x is the range that includes $90\%$ of the data (excluding $5\%$ at either end).  The resulting overlap regions for each parameter are demarcated in gray in Fig.~\ref{fig:findcrit_all}, and the upper left corner of each panel shows the two measures $\Delta$N/N and $\Delta$x/x. These quantitative measures confirm what was seen by eye, namely, that $M_*/r_{\rm e}$ gives the smallest values of both $\Delta{\rm N/N}$ and $\Delta$x/x among all the mass-radius combinations. Note that the relative extent of $M_B$ is $\Delta$x/x=0.971; this simply means that the overlap region is almost equal to the entire range that encompasses $90\%$ of the data, again agreeing with our previous qualitative assessment. Furthermore, we find that S\'ersic index performs considerably better than even $M_*/r_{\rm e}$ in minimizing both $\Delta{\rm N/N}$ and $\Delta$x/x.

We also point out the extremely tight relation that is produced when plotting color vs.~$M_*/r_{\rm e}$ or $M_*$ {\it for blue-cloud galaxies alone} (Fig.~\ref{fig:findcrit_all}b and \ref{fig:findcrit_all}c). This has been pointed out before and is referred to as the `Main Sequence' of star formation \citep{noeske07}. Previous work on quenching has focused on the relationship between red sequence vs.~blue cloud galaxies and not so much on the properties of galaxies within the blue cloud itself. However, the tightness of the relation between $M_*/r_{\rm e}$ (or $M_*$) and color for star-forming galaxies alone could be an important clue to the physics of quenching, and we return to this point in \S\ref{sec:Discussion}.

 \begin{figure*}[t!]
 \centering
\includegraphics[scale=.75,angle=90]{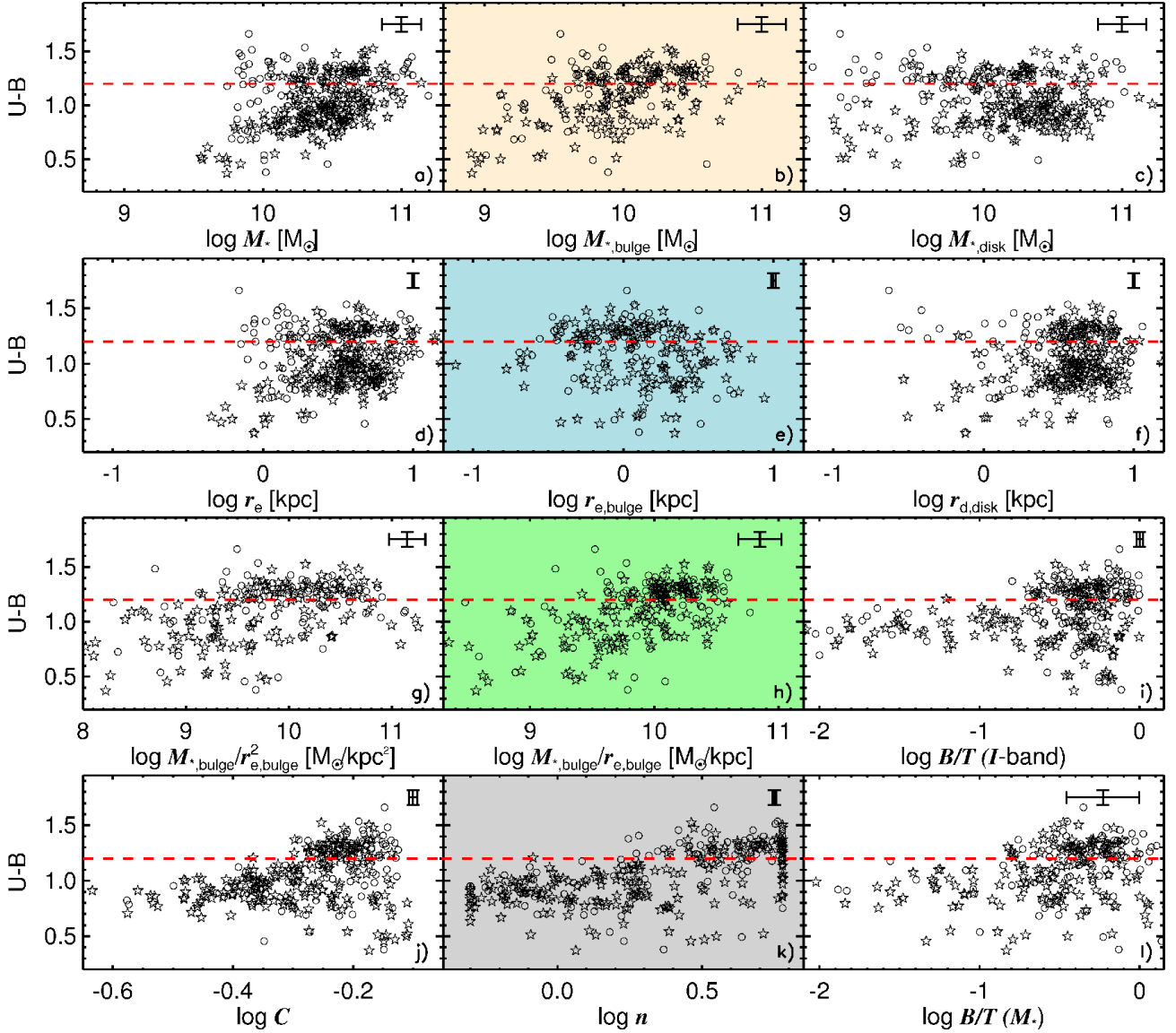}
\caption{Galaxies in the overlap region of $M_*/r_{\rm e}$ (containing 316 galaxies) from Fig.~\ref{fig:findcrit_all}c are isolated to examine possible second-parameter color correlations with both global and disk/bulge properties. $U-B$ rest-frame color is plotted against: a) global stellar mass $M_*$, b) bulge stellar mass $M_{*,\rm bulge}$, c) disk stellar mass $M_{*,\rm disk}$, d) global semimajor axis effective radius $r_{\rm e}$, e) bulge semimajor axis effective radius $r_{\rm e,bulge}$, f) disk semimjaor axis exponential scale length $r_{\rm d,disk}$, g) $M_{\rm *,bulge}/r_{\rm e,bulge}^2$, h) $M_{\rm *,bulge}/r_{\rm e,bulge}$, i) bulge-to-total ratio $B/T$ in the $I$-band, j) concentration $C$, k) global S\'ersic index $n$, and l) $B/T$ based on stellar mass $M_*$. The 1-$\sigma$ error bars are displayed in the upper right of each panel. The red horizontal line in each panel represents the division between red and blue galaxies. Global properties hardly differ between red and blue galaxies (white panels), but bulge properties, disk scale length, and $n$ show key differences (colored panels and panel c). Specifically, red sequence bulges are physically smaller, yet more massive, than those of the blue cloud. Moreover, red sequence disks are less massive than their blue cloud counterparts. These trends cannot be produced by simple disk fading but require a concentration of inner stellar mass.
 \label{fig:second_aegismeet}}
 \end{figure*}

\subsection{Properties of Galaxies in the $M_*/r_{\rm e}$ Overlap Region} \label{sub:m/r}

We have shown that our AEGIS data duplicate previous findings showing that $M_*/r_{\rm e}$ and $n$ correlate strongly with quenching, but we have also shown that neither parameter alone is close to being a perfect predictor of it.  In this section and the next, we take a further look at the properties of galaxies in the overlap region and outliers to find out whether multiple parameters can be used in concert to predict quenching, and whether this interplay sheds light on the physical processes involved.

Fig.~\ref{fig:second_aegismeet} investigates residual trends within the $M_*/r_{\rm e}$ overlap region of Fig.~\ref{fig:findcrit_all}c by plotting color versus various structural parameters for overlap galaxies alone.  In exploring this slice of $M_*/r_{\rm e}$, we are assuming a general evolutionary sequence such that the blue galaxies evolve into the red galaxies within this overlap region. However, this assumption is not without proof. \cite{bell04} and \cite{faber07} have shown that the red sequence has increased by $\sim2$ while the blue cloud has remained relatively stable from $z\sim1$ to $z=0$. Moreover, \cite{hopkins10b} showed that $\sim65-80\%$ of the observed mass density of bulge-dominated galaxies formed since $z\sim1$. Thus our redshift range ($0.5\le z < 0.8$) peers directly into the epoch when the majority of red sequence galaxies are being formed. We choose $M_*/r_{\rm e}$ as the base parameter because it is the tightest $M_*-r_{\rm e}$ combination in Fig.~\ref{fig:findcrit_all}, but similar results are obtained when $M_*$ is used. 

Panels a, d, j, and k of Fig.~\ref{fig:second_aegismeet} plot integrated quantities, while the remaining panels introduce structural parameters (e.g., $M_*$ and $r_{\rm e}$) for bulges and disks separately.  Among the integrated properties, virtually no trend is seen in either stellar mass (panel a) or $r_{\rm e}$ (panel d), but a mild ``step function'' is seen with S\'ersic index in that $n$ is significantly higher for quenched galaxies (panel k, gray).   Concentration, $C$, shows similar behavior, albeit less cleanly (panel j). A similar trend with $n$ was seen for {\it all} galaxies (Fig.~\ref{fig:findcrit_all}e), but it is important to see the same effect for overlap galaxies alone.  This establishes beyond doubt that $M_*/r_{\rm e}$ alone does not fully encapsulate the processes needed to quench star formation.  A possible interpretation is that some galaxies are ``ripe'' for quenching based on $M_*/r_{\rm e}$ and that a second process, which drives galaxies to high $n$, ultimately quenches star formation.  We return to this idea later in \S\ref{sec:Discussion}.

The remaining panels in Fig.~\ref{fig:second_aegismeet} focus on the properties of bulge and disk components separately (the subcomponent sample; see \S\ref{sub:gim2d}). These parameters are derived from GIM2D photometric fits and $M_*/L_{B}$ values from the $V - I$ colors of bulges and disks separately (\S\ref{sub:mass}), for which high-resolution two-color {\it HST} imaging is required.  The striking result from the subcomponent panels in Fig.~\ref{fig:second_aegismeet} is the marked differences in {\it disk} mass, {\it bulge} mass, and {\it bulge} effective radius between blue and red overlap galaxies (panels b, c, and e). The disks of red sequence galaxies are less massive by about 0.2 dex than the blue cloud galaxies (panel c) while the bulges of red sequence galaxies are more massive by about 0.3 dex than the blue cloud galaxies (panel b, in light tan). At the same time, disk sizes remain constant but bulge sizes decrease by about 0.3 dex as blue cloud galaxies transition to the red sequence (panel e, in blue). These differences between red sequence bulges and blue cloud bulges amount to a difference of 0.6 dex in bulge $M_{*,\rm bulge}/r_{\rm e, bulge}$ (panel h, in green). A similar trend is seen in $B/T$ ratios, but it is weaker due to the large spread of blue galaxies. 

These trends collectively demonstrate a real structural difference between the {\it inner stellar mass distributions} of quenched vs. star-forming galaxies, and furthermore, that this difference exists even within a narrow range of $M_*/r_{\rm e}$.  Higher central stellar mass densities in quenched galaxies have been inferred in previous work from higher integrated S\'ersic values \citep{weiner05, bundy10}, but photometric data by themselves do not rule out a simple fading picture in which disks decline in brightness, permitting an underlying high-$n$ bulge to emerge \cite[see also][for discussion of $b/a$]{holden12}.  Actual stellar masses for bulges and disks separately are needed to rule out fading.  An important new insight from our work is that evolution to the red sequence appears to be accompanied by a {\it significant rearrangement of inner stellar mass} in which existing stars move to the centers of galaxies, and/or new stars are formed there.  We discuss processes whereby that might happen in \S\ref{sec:Discussion}.

\begin{figure*}[t!]
 \centering
\includegraphics[scale=.6]{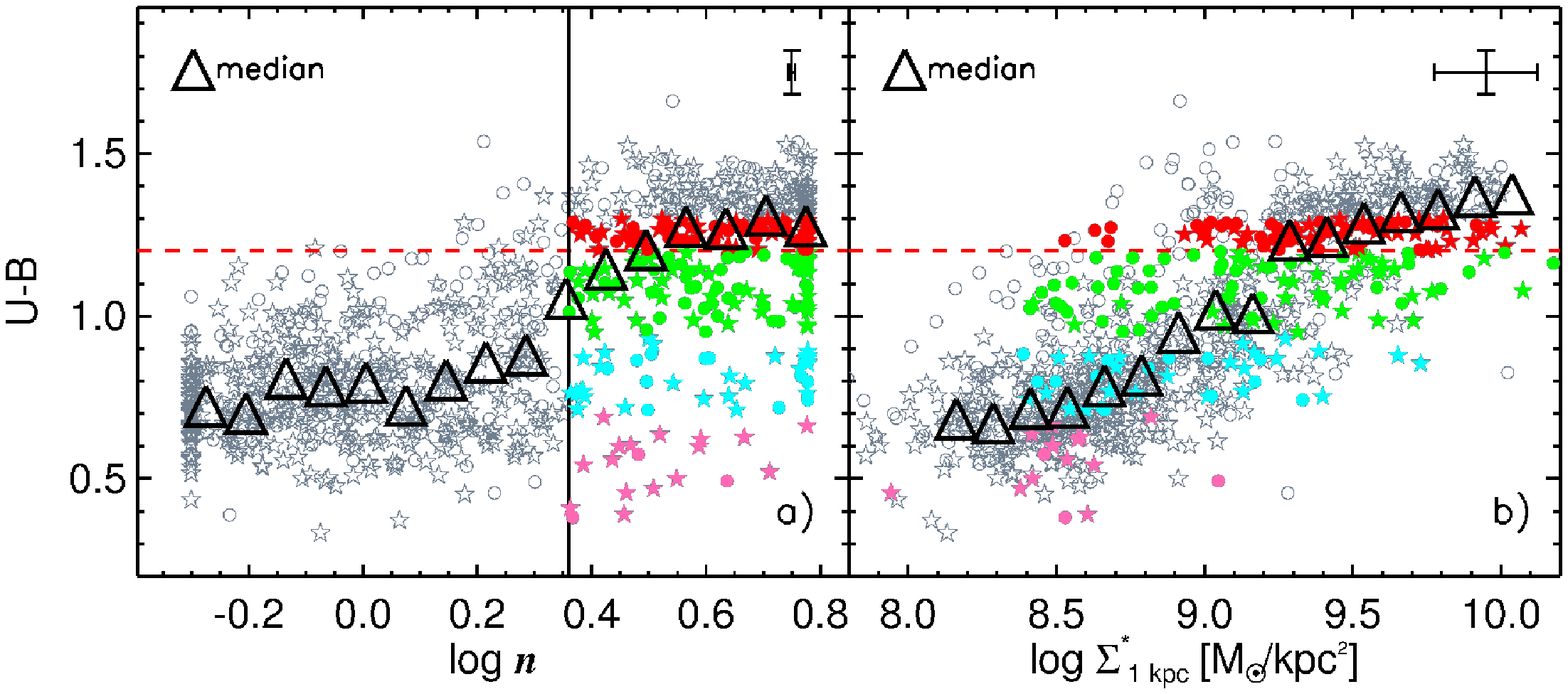}
\caption{a) The $U-B$ vs.~$n$ diagram is reproduced from Fig.~\ref{fig:findcrit_all}e with median values of color in each $n$ bin plotted as open black triangles. The red horizontal line represents the division between red and blue galaxies, while the black vertical solid line at $\log~n=0.36$ represents the point at which the median is half way between the red and blue values.  The medians show a step-like behavior in $U-B$ vs.~$n$, suggestive of a physical threshold in $n$.  Exceptions to the step function are the blue ``outlier'' galaxies in the lower right.  These are highlighted in strips of color for further discussion in \S\ref{sub:outliers} and Fig.~\ref{fig:n_strip}: pink points encompass $U-B<0.70$, cyan points encompass $0.70 \le U-B <0.95$, and green points encompass $0.95 \le U-B < 1.20$. Bluish red-sequence galaxies are highlighted for comparison and lie within $1.20\le U-B<1.30$. Images of these outliers are shown in Fig.~\ref{fig:n_strip}.  Roughly 40\% of the $\log~n>0.36$ galaxies are outliers.  b) $U-B$ color vs. inner stellar mass surface density $\Sigma^*_{\rm 1 kpc}$. Note that some galaxies do not have $\Sigma^*_{\rm 1 kpc}$ measurements due to insufficient signal-to-noise. Most outliers now fall into line, suggesting that inner stellar mass surface density is a cleaner predictor of quenching than $n$.
  \label{fig:nvsub}}
\end{figure*}

\subsection{S\'ersic Index and Inner Surface Density} \label{sub:sersic}

\subsubsection{A threshold in $n$?} \label{sub:threshold}

The previous section considered $M_*/r_{\rm e}$ as the main quenching parameter and looked at scatter around it to discover secondary effects.  In this section, we use a similar tack but focus on S\'ersic index $n$.  An enlarged version of the color-$n$ plot is shown in Fig.~\ref{fig:nvsub}a, which indicates median $U-B$ in bins of $n$.  The medians illustrate the striking step-like behavior previously mentioned.  Defining the riser of the step to be where the medians are half-way between their red and blue values, we place this point at $\log~n=0.36$, which is marked with the vertical black line. This value corresponds to $n = 2.3$, which is similar to the value of $n$ often used to distinguished quenched (or early-type) galaxies from star-forming ones locally \citep [e.g.,][]{shen03, bell04a, schiminovich07, drory07}.

The medians also emphasize how {\it flat} the color trends are above and below the threshold $n$ value.  Evidently, in the extreme high-$n$ and low-$n$ regimes, star formation history is independent of $n$.  This differs markedly from the behavior of $M_*/r_{\rm e}$; Fig.~\ref{fig:findcrit_all}c shows a strong correlation between $M_*/r_{\rm e}$ and color for star-forming galaxies {\it below} the overlap region.   

The lack of importance of $n$ above and below the threshold is further emphasized by the large color scatter in both of these regimes.  This scatter is, however, of two types.  At low $n$, there is a rather uniform spread in color, i.e., specific star-formation rate can assume any value within a large range, and $n$ does not predict what SSFR will be.  At high $n$, $n$ predicts color much better, i.e., the color distribution is strongly peaked toward red (quenched) colors, but a significant tail of outliers with blue colors exists (colored points, except the red in Fig.~\ref{fig:nvsub}).  The existence of these outliers was seen at both low and high redshifts by \cite{bell08} and \cite{bell12}, who expressed the role of $n$ in quenching as ``necessary but not sufficient'', i.e., all quenched galaxies have high $n$, but not all high-$n$ galaxies are quenched.  We see the same thing.

\begin{figure*}[t!]
 \centering
\includegraphics[scale=.70,angle=90]{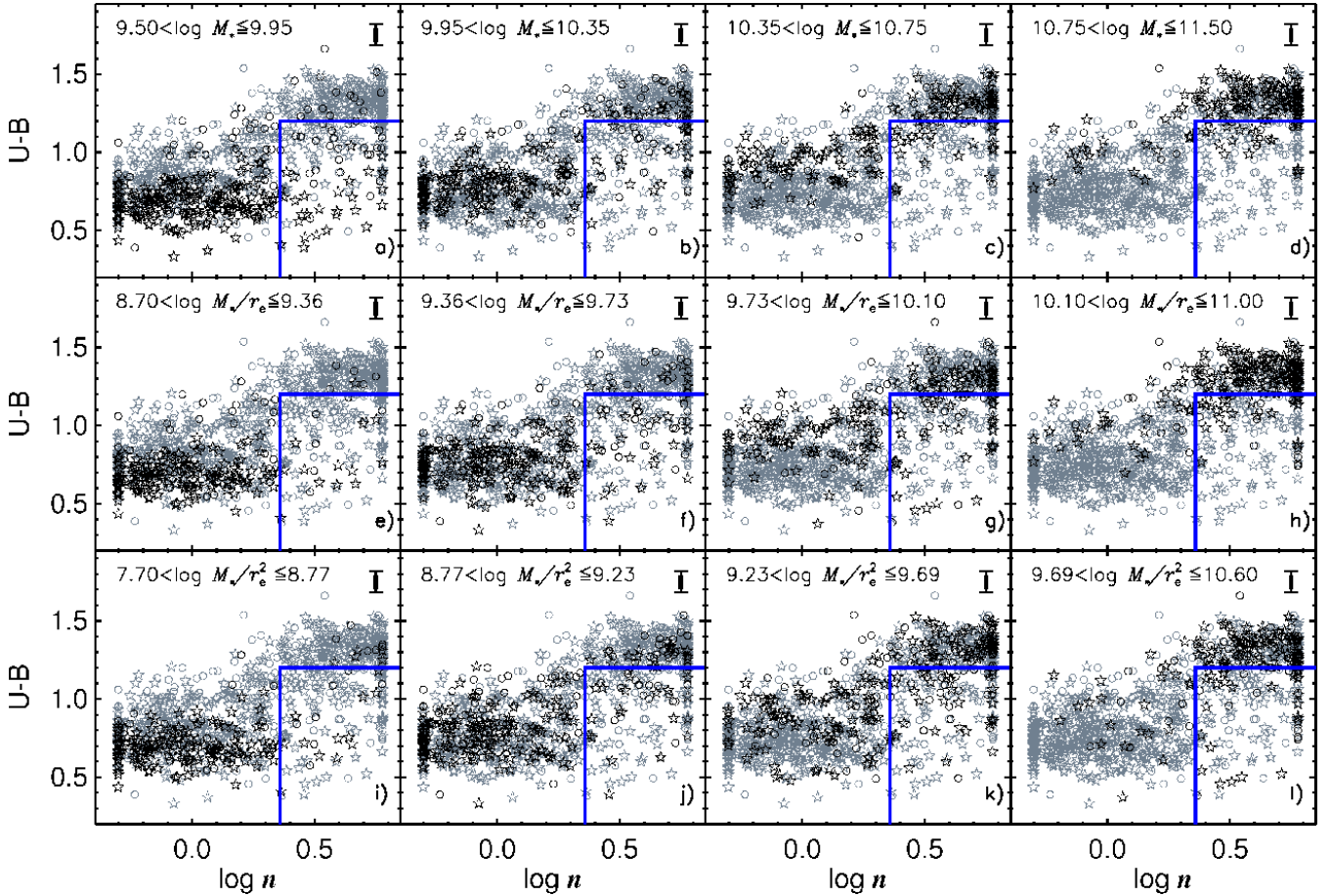}
\caption{The $U-B$ vs.~$n$ diagram from Fig.~\ref{fig:nvsub}a is replotted to separate the roles of $M_*$, $M_*/r_{\rm e}$, and $M_*/r_{\rm e}^2$ from $n$ in driving color evolution.  Galaxies are divided into bins of $M_*/r_{\rm e}^p$ and plotted as the black points.  Bin boundaries are shown in the upper left of each panel.  The rectangles outlined in blue indicate the outlier region from Fig.~\ref{fig:nvsub}a.  The panels demonstrate that the step-function-like behavior near $\log~ n=0.36~(n=2.3)$ seen in Fig.~\ref{fig:nvsub}a is replicated separately in each range of $M_*/r_{\rm e}^p$ independent of $p$.  Color rises rapidly near $\log~ n=0.36~(n=2.3)$, but there is little impact of $n$ on color above and below this value.  This shows that the trends in color with $M_*$ and $M_*/r_{\rm e}$ in the blue cloud (Fig.~\ref{fig:findcrit_all}b and \ref{fig:findcrit_all}c) are not caused by a hidden dependence on $n$. 
 \label{fig:nvsub_mass}}
 \end{figure*}

Unlike Fig.~\ref{fig:findcrit_all}c (which plotted color vs.~$M_*/r_{\rm e}$), there is no interval in $n$ where the color scatter is markedly larger than elsewhere (Fig.~\ref{fig:nvsub}a), and thus no impetus to search for a second parameter within a {\it narrow region} of $n$.  To investigate the scatter, we have replotted Fig.~\ref{fig:nvsub}, this time highlighting galaxies within narrow bins of various second parameters. The results are shown in Fig.~\ref{fig:nvsub_mass}, where galaxies are divided into bins of stellar mass (top row), $M_*/r_{\rm e}$ (middle row), and $M_*/r_{\rm e}^2$ (bottom row), collectively termed $M_*/r_{\rm e}^p$.  The outlier region from Fig.~\ref{fig:nvsub}a is outlined in blue. In each row, the behavior is the same. Galaxies with low values of $M_*/r_{\rm e}^p$ are seen to be mainly blue.  A few leak into the high-$n$ ``outlier'' regime, but their blue colors always agree with other galaxies in the same parameter bin, i.e., their star formation rates are not depressed by having high $n$.  As $M_*/r_{\rm e}^p$ increases, the mean color of low-$n$ galaxies becomes redder while the number of outliers remains relatively constant.  Again, the colors of the outliers still agree with the average color of all galaxies in the same $M_*/r_{\rm e}^p$ bin.  At the highest values of $M_*/r_{\rm e}^p$, virtually all galaxies are quenched and the fraction of outliers is negligible. Two points are clear: dividing galaxies into bins has not destroyed the basic step-like nature of the behavior in that galaxies within each bin still trend smoothly but sharply (apart from outliers) from their ``native'' star-forming state to a quenched state.  The second point is that all trends with color at low $n$ remain flat within each bin of $M_*/r_{\rm e}^p$.  This shows that the strong trend of color vs.~$M_*$ or $M_*/r_{\rm e}$ within the blue cloud (Fig.~\ref{fig:findcrit_all}b and ~\ref{fig:findcrit_all}c) is not caused by some hidden dependence on $n$.

To summarize, we have reproduced findings by previous authors that indicate that high $n$ typically predicts a quenched galaxy, and we have set the half-power point between blue and red galaxies at $n_{crit} = 2.3$.  This value is very near the SDSS value, implying no large evolution in $n_{crit}$ from $z\sim0.65$ down to $z \sim 0$.  The rise in color near the critical value is sharp, while above and below this value there is no trend in color with $n$, even within narrow slices of $M_*/r_{\rm e}^p$.  At high $n$, most galaxies are quenched with red colors, but a non-negligible fraction of objects is blue despite having high $n$. We turn to the nature of these outliers next.

\subsubsection{Outliers} \label{sub:outliers}

Although $n$ acts like a threshold for the vast majority of galaxies, there are obvious exceptions, namely the blue, high-$n$ $(\log n \ge 0.36)$ ``outliers'' highlighted in Fig.~\ref{fig:nvsub}a and elsewhere.  Understanding these objects is clearly crucial for unraveling the quenching mechanism -- why are they blue when their photometric structure resembles that of quenched objects?   We have highlighted 151 outliers in Fig.~\ref{fig:nvsub} using color to indicate their $U-B$ ranges; they make up $\approx40\%$ of all $n>2.3$ galaxies (the red points immediately above the red horizontal dashed line at $U-B=1.20$ are not outliers, they are quiescent red sequence galaxies shown for comparison).

Several possibilities come to mind to explain these objects.  One possibility is that they are artifacts due to the presence of bright point-like AGNs.  Adding a blue AGN to a normal star-forming galaxy would make the global colors bluer and increase $n$ (and concentration) \citep{pierce10}.  To pursue this, we have cross-matched the outliers to two AGN samples selected using X-ray and optical line-emission data.  The AEGIS region is covered by a deep 800 ksec $Chandra$ X-ray mosaic \citep{laird09}.  We find that only 11 of the outliers have X-ray luminosities above $10^{42} ~ \rm erg ~s^{-1}$, or $7\%$.  We have also used an optical method for selecting AGN based on a modified version of the classical ``BPT'' diagram \citep{bpt81} that plots $\OIII$/$\Hb$ versus $U-B$ rather than  $\NII$/$\Ha$ \citep{yan11}.  This adds only 14, bringing the total to 21 AGN, or $14\%$ of the outliers.  Thus, it seems that the vast majority of these objects are unlikely to be AGN hosts.

\begin{figure*}[t]
 \centering
\includegraphics[scale=.7]{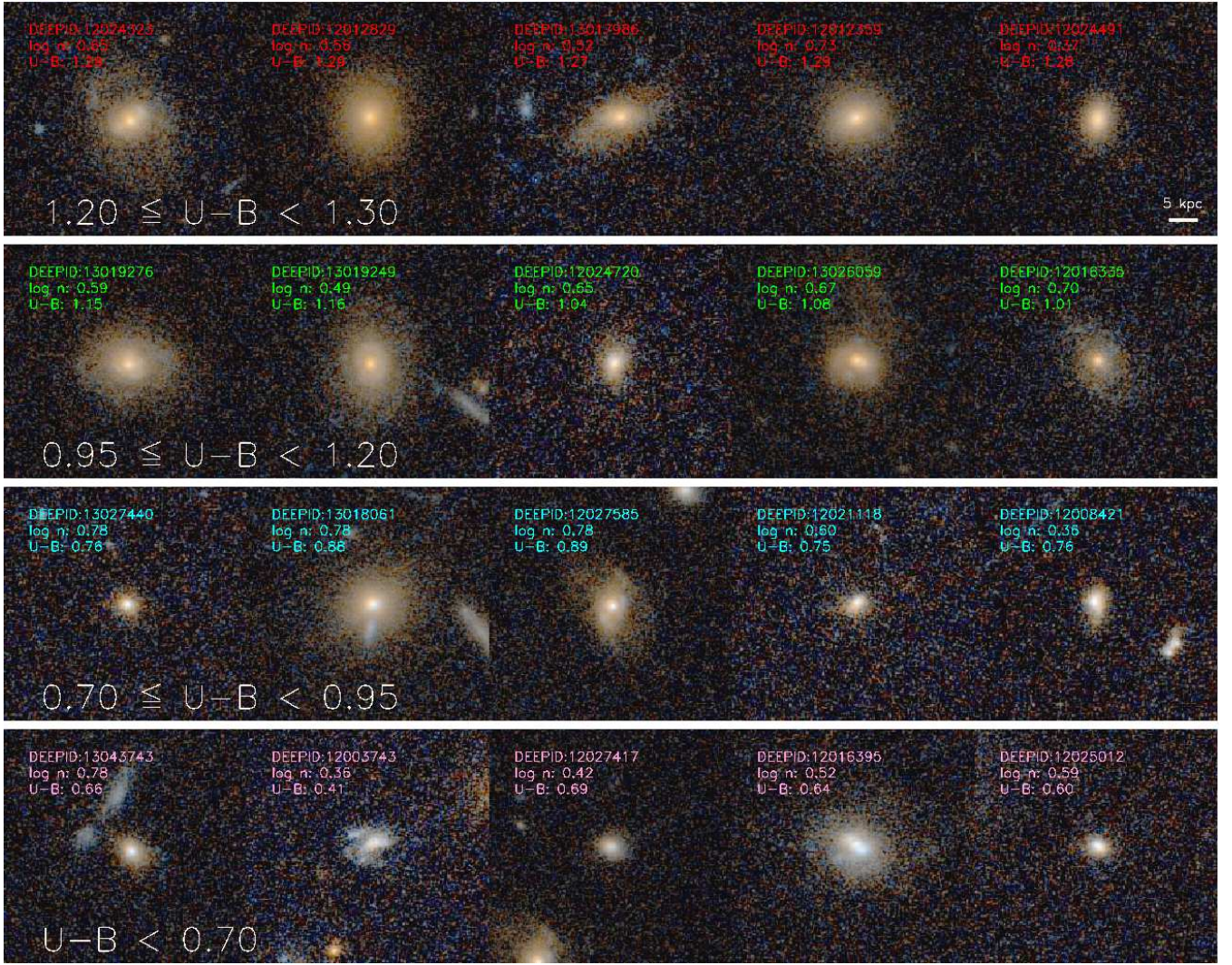}
\caption{$HST$/ACS $V$ and $I$ color images for a selection of high-$n$ outlier galaxies from Fig.~\ref{fig:nvsub}a.  Objects are arranged in rows according to $U-B$ color with bluer galaxies at the bottom.  The bottom row corresponds to the pink points in Fig.~\ref{fig:nvsub}a with $U-B<0.70$; the next row up corresponds to cyan points with $0.70 \le U-B <0.95$; and the following row is the green points with $0.95 \le U-B < 1.20$. The top row contains red sequence galaxies shown for comparison; they are represented by the red points in Fig.~\ref{fig:nvsub}a and have $1.20\le U-B<1.30$. A scale of 5 kpc at the average redshift ($z\approx 0.68$) of our sample is shown as a reference in the upper right picture. Many of these blue high-$n$ objects seem to have S\'ersic indices perturbed by small size, irregular structure, central starbursts, and/or AGN and are cured as outliers if inner stellar surface density $\Sigma^*_{\rm 1 kpc}$ is used instead of $n$, as shown in Fig.~\ref{fig:nvsub}b.  
 \label{fig:n_strip}}
 \end{figure*}

Another possibility is a bright blue clump of recently formed stars at the centers of the outliers, which might skew the colors and S\'ersic indices as an AGN would. We would like to remind the reader that the single S\'ersic GIM2D model does not fit for any substructure, such as clumps. Hence asymmetric structures may affect the $n$ measurements. To explore this, we constructed $V$ and $I$ color images of all outliers using the $HST$/ACS data and inspected them; a montage of 20 galaxies is shown in Fig.~\ref{fig:n_strip}, where each row represents a different range of $U-B$ color according to the color-coding in Fig.~\ref{fig:nvsub}a.  The bluest outliers are in the bottom row.  These tend to be lumpy, asymmetric, and/or small -- their fitted S\'ersic values are somewhat questionable.  Moving up one row to the cyan points, we find a mix of genuinely concentrated galaxies plus more small galaxies like the ones in the previous row. The third row contains larger objects of intermediate color but with convincingly high concentrations.  Finally, we show a sample of red sequence galaxies in the top row as a comparison; they are all red and highly concentrated spheroids. 

To summarize, the blue, high-$n$ outliers are a mix of different types. Some may have doubtful S\'ersic indices, being small or with off-centered clumps of star formation or AGN, but a fair fraction seem to be genuinely blue yet high-$n$ galaxies.  These genuine exceptions tend to be located at intermediate values of $U-B$.  The existence of such outliers has been noticed before.  An important class of candidates is {\it poststarburst galaxies} \citep[e.g.,][]{dressler83, couch87, poggianti99, goto03}.  These objects possess blue colors and strong Balmer absorption yet weak H$\alpha$, signifying recent quenching, and their S\'ersic indices are high \citep{quintero04, yang08}. A second class of objects is the {\it blue spheroidal galaxies}; like poststarbursts, they have smooth, centrally concentrated, elliptical-like profiles but they are different in having active star formation \citep{menanteau01, im01, koo05, schawinski09, kannappan09}. Their masses tend to be small \citep{im01}, and there appear to be several objects in our outlier population that fit this description in the bottom row of Fig.~\ref{fig:n_strip}. 

A quick calculation of the percentage of outliers within a volume-limited SDSS sample at $z\sim0$ shows that it has dropped from $\approx40\%$ for our sample at $z\sim0.65$ to $\approx10\%$ at $z\sim0$. This difference seems consistent with the higher levels of gas at higher redshifts, which could give rise to more clumpy star formation asymmetrically distributed throughout the galaxy, skewing the S\'ersic values.  

For completeness, in passing we also mention satellite processes. Processes such as ram pressure stripping \citep{gunn72} and strangulation \citep{larson80} do not by themselves change $n$. However, other satellite mechanisms involving tidal interactions (such as ``harassment'' e.g., \citealt{moore96}) may induce a morphological transformation.  If satellite galaxies are first harassed, they might attain a high $n$ while still forming stars. While we do not expect most of the DEEP2 galaxies to be satellites (see introduction), a more thorough investigation is needed to confirm this.
  
\subsubsection{Inner Surface Density: An Improvement on n?} \label{sub:surfacedensity}

From the above, it is clear that in $n$ we have found a remarkable, but still imperfect, structural predictor of quenching.  The main criticism of $n$ is the presence of outliers; if they could be removed, the correlation, and hence the prediction, would be nearly exact.  

There are two obvious weaknesses with $n$. First, it is based on light, not stellar mass, and hence is subject to the vagaries of star formation history and dust.  Second, it relies on a fit to the entire light profile and is thus at least partially dependent on the outer light distribution, which may be disturbed or irregular.  In contrast, trends discovered using bulge properties in Fig.~\ref{fig:second_aegismeet} hint that the structure of the {\it inner regions} of galaxies is more important at predicting quiescence.

Accordingly, we introduce a new parameter that attempts to remove both of these weaknesses, namely the {\it stellar mass surface density within 1 kpc}.  This is defined as $\Sigma_{\rm 1 kpc}^* \equiv M^*_{\rm 1 kpc}/\pi r_{\rm 1 kpc}^2$, where $r_{\rm 1 kpc} \sim 1$ kpc\footnote{Note that here we include the $\pi$, so these are physical surface densities.}.  The actual diameter used is 12 pixels $(0.36 \arcsec)$, which is set by the smallest radius that our {\it HST} images can conveniently resolve; it spans a radius of $1.08-1.35$ kpc at our redshifts.  The quantities $I$ and $V-I$ are measured within this aperture, and $L_B$, $M_*/L_B$, and $M_*$ are estimated using Eqns.~\ref{eqn:mb},~\ref{eqn:kcorrect}, and \ref{eqn:masslight}.  

The quantity $\Sigma_{\rm 1 kpc}^*$ was already included for completeness in Fig.~\ref{fig:findcrit_all} (panel f), where its performance is seen to be mixed.  It seems to predict color less well than the global quantities $M_*$ and $M_*/r_{\rm e}$ in the blue cloud, but it does much better than $n$ in removing outliers.  This is even better illustrated in Figs.~\ref{fig:nvsub}a and \ref{fig:nvsub}b, which compare outliers directly. Using $\Sigma_{\rm 1 kpc}^*$, almost all the pink and cyan points have receded back into the blue cloud, and only a few green outliers remain. This suggests that the outliers in $n$ are largely artifacts caused by poor GIM2D fits\footnote{GIM2D only models a galaxy into either a bulge+disk model or a single S\'ersic model. More complex structures like spiral arms, bars, and clumps are not modeled. Thus a galaxy with these features are potentially ill-fit.} and that using a more secure quantity like inner mass surface density can remove them.  It is still true that using $\Sigma_{\rm 1 kpc}^*$ by itself is not perfect and that $n$ somewhat outperforms it on the overlap criteria seen in Figs.~\ref{fig:findcrit_all}e and \ref{fig:findcrit_all}f. That a genuine spread exists in $\Sigma_{\rm 1 kpc}^*$ is confirmed by ongoing work with higher-S/N SDSS data (Fang et al., in prep), which however reveals some additional striking regularities.  Our point for now is that using $\Sigma_{\rm 1 kpc}^*$ removes the large number of outliers that are present with $n$.  Furthermore, the definition of $\Sigma_{\rm 1 kpc}^*$ as an {\it inner} mass density directs our attention even more strongly to the fact that it is conditions {\it near the center of the galaxy} that drive quenching.

\section{Discussion} \label{sec:Discussion}

In this paper, we have found that the most discriminating parameter of quiescence, according to the measures of the overlap region, is the S\'ersic index, and that a plot of color vs. S\'ersic index shows a step-like behavior near $n=2.3$, suggestive of some sort of genuine quiescence threshold. About $40\%$ of the $n>2.3$ galaxies, however, are ``outliers'' that fall outside this behavior. These outliers have central mass densities much lower than those of the red sequence and fully consistent with those of the blue cloud. In other words, under this new parameter, $\Sigma_{\rm 1 kpc}^*$, the outliers now fall in line, suggesting that a galaxy's central structure may be even more physically related to quiescence than S\'ersic index.

Both $n$ and $\Sigma_{\rm 1 kpc}^*$ corroborate our second major finding, that most blue cloud galaxies at the observed epoch cannot simply fade onto the red sequence. We have shown through stellar mass measurements of bulges and disks that the red sequence galaxies have absolutely higher bulge {\it mass} concentrations, i.e., that the jump in $n$ is not due simply to the fading of disks (see \S \ref{sub:m/r}). The central mass densities extend this conclusion to the very centers of galaxies. In other words, the transition to the red sequence involves a significant restructuring of a blue cloud galaxy's {\it innermost stellar density profile}. 

Below the critical value of $n=2.3$, however, S\'ersic index shows little to no correlation with star formation, and color is more closely correlated with $M_*/r_{\rm e}$ (or perhaps with $M_*$, see Fig.~\ref{fig:findcrit_all}). 

This two-pronged behavior suggests that the star-formation history of a galaxy is shaped by two separate factors at different stages.  While the object is still star-forming (in the blue cloud), its star-formation rate depends on {\it global} parameters, like $M_*/r_{\rm e}$ or $M_*$. Then, a major internal mass reorganization occurs, a dominant bulge forms, and star formation stops. In the following discussion, we compare these results to the predictions of the standard merger model for bulge-building and quenching and find reasonable agreement, but also several issues. To alleviate these issues, we also consider other models, specifically, violent disk instability \citep{noguchi99, elmegreen08, dekel09}, secular evolution \citep{kk04}, morphological quenching \citep{martig09}, and halo quenching \citep{silk77, rees&ostriker77, blumenthal84, birnboim03, keres05, dekel06, cattaneo06}. We end with a brief discussion on an unresolved concern.

\subsection{Merger Model} \label{sub:mergers} 

Major mergers\footnote{According to \cite{hopkins10}, major mergers dominate the formation and assembly of $\sim L_*$ bulges and the total spheroid mass density. Thus, we only consider major mergers in this discussion. However, it should be noted that minor mergers can create bulges \citep{bournaud07} and do contribute a non-negligible amount ($\sim 30\%$) to the total spheroid mass density \citep{hopkins10}.} have been linked to the formation of spheroids since \cite{toomre72}, with considerable work in the years since (see \citealt{hopkins09a} and references therein). The process of bulge formation in classical merger models occurs through both the violent relaxation of pre-existing stars to the center and a gaseous dissipation-induced nuclear starburst \citep{hopkins09a}. 

Comparison of the bulge-dominated products of these simulations to observed early-type galaxies shows good agreement. The S\'ersic indices of the merger products from \cite{hopkins08} are high, $n \gs 2.5$. Similarly, \cite{naab06} showed that their disk mergers (with bulges) created galaxies with $3<n<4$ \footnote{\cite{naab06} conducted a collisionless simulation that does not include gas, and thus does not model a nuclear starburst component. That is why their pure disk-disk mergers only have $n\approx1.5$, because they lack the high central surface brightness typical of dissipative gas-rich mergers.}. Both of these works produce spheroids with S\'ersic indices in the range of our red sequence spheroids and of other observations \citep[e.g.,][]{kk04, drory07, fisherdrory08}. 

The properties of quenching induced by mergers are also consistent with our data. The merger model predicts that quenching occurs through the nuclear starburst \citep{mihos94}, in which a large portion of the gas\footnote{In the following discussion, gas is assumed to be cold gas unless otherwise stated} is exhausted (depending on gas fraction; see \citealt{springel&hernquist05, okamoto05, robertson06, cox08, governato07, governato09}), and through subsequent feedback, from supernovae and/or AGN \citep{springel05, murray05, ciotti09}. Note that both sources of quenching originate {\it from the center of the galaxy}, suggesting that the conditions at the center may correlate better with the quenching state than global properties. This is what we find. Furthermore, the central surface mass densities of the simulated spheroids from \cite{hopkins09b} match our observations of the red sequence spheroids in Fig.~\ref{fig:nvsub}b -- with values of $\log ~\Sigma = 9 - 10~ \rm{M_{\odot} kpc^{-2}}$ at 1 kpc.

A further success of the merger hypothesis is the good match between it and the stellar mass range where bulge-building is observed to occur.  A major point is that the efficiency of bulge-building from major mergers is expected to be highly dependent on the {\it pre-merger gas fraction}, such that decreasing gas content increases the potential to form bulges \citep{springel&hernquist05, robertson06, hopkins09a}. This dependence is consistent with the assumption that gas content gradually falls as galaxies age in the blue cloud, making them ultimately ripe for spheroid formation via mergers. Because more massive galaxies exhaust their gas quicker due to the phenomenon known as ``downsizing'' \citep[e.g.,][]{cowie96}, there is a strong color-mass relation in the blue cloud, meaning that the reddest blue cloud galaxies at any epoch have the least amount of gas.  According to the merger model, this means that they are also on the threshold of forming bulges. 

Evidence for this hypothesis is strong in Fig.~\ref{fig:findcrit_all}b, which shows a remarkably tight correlation between stellar mass and color in the blue cloud in the sense that more massive galaxies are the reddest.  Further data come from \cite{catinella10} and \cite{saintonge11}, who presented $\rm H~I$ and CO data in the GASS and COLD GASS survey, respectively. These works illustrate that the average atomic and molecular gas fraction of galaxies do decrease with increasing stellar mass and increasing NUV-r color. Although these surveys do not fully sample the blue cloud (these surveys only observe $M_*>10^{10} ~\rm M_{\odot}$ galaxies), extrapolating these seemingly linear trends to lower masses indicate that total gas fraction does indeed decrease with mass along the blue cloud. Theoretically, \cite{hopkins10} showed that the most effective bulge-building major mergers are clustered around $\log~M_* \sim 10.5 ~\rm M_{\odot}$ at $z\sim1$, which is in the center of overlap region of $M_*$. This behavior is due in part to exactly the same reason, namely, decreasing gas content as galaxies age within the blue cloud.  In a general way, then, theory predicts that galaxy colors and gas contents should both age within the blue cloud, causing galaxies to become more prone to bulge-building mergers at higher mass and low gas level, and these trends broadly agree with the observations.

In detail, however, the data indicate that $M_*/r_{\rm e}$ is a better predictor of quenching than $M_*$ alone (cf. overlap regions in Fig.~\ref{fig:findcrit_all}b vs.~\ref{fig:findcrit_all}c). This may be because $M_*/r_{\rm e}$ is related to velocity dispersion (\citealt{franx08}; see footnote 26), which, based on a new study by \cite{wake12b}, is the galaxy property most closely related to halo mass. This finding could then be a manifestation of the dependence of quenching on a critical halo mass.  Alternatively, it may reflect the fact that radii are shrinking as stellar mass builds up in the centers of quenching galaxies and thus reflects a property of the galaxies themselves rather than of their halos. We elaborate further on these thoughts in \S\ref{sub:halomass}.

Although there are some aspects of the merger model that match our data, general agreement upon the validity of this model has not been reached. One controversial issue is whether major mergers can actually quench galaxies. There have been various works that support this idea \citep[e.g.,][]{schawinski07b, alexander10, cano-diaz11, farrah11}. In particular, \cite{cano-diaz11} obtained VLT-SINFONI integral-field spectroscopy for one quasar at $z=2.4$ and showed a suppression of narrow H$\alpha$ emission, a tracer of star formation, in the region with the highest outflow velocity and highest velocity dispersion. But this is only one example and does not erase the contradicting evidence others have offered. For example, using a sample of X-ray and post-starburst galaxies from SDSS and DEEP2 at $0.2 < z < 0.8$, \cite{coil11} found winds with velocities that are insufficient to shut down star formation, indicating that the presence of an AGN does not produce faster winds nor does it seem to play a major role in quiescence. And \cite{ammons11} fail to find any correlation between host galaxy color and X-ray hardness ratio among $z = 0.5-1.5$ galaxies, as might be expected from the blowout model.

Another important concern is whether there are enough mergers to account for the bulge density in the universe. Studies that have measured the galaxy merger rate often present different results \citep[e.g.,][]{bell06, conselice06, lin08, lotz08b, bundy09, xu11, bluck11}. \cite{lotz11} addressed the issue of disparate observational merger rates; they found that the major reason for these differences is the assumed timescale in which a merger is observable. Using a suite of hydrodynamic merger simulations, they constrained the observable timescales of three common merger rate estimators -- close galaxy pairs, G/$M_{20}$, and asymmetry -- and found that if a physically motivated average observability timescale was adopted to calculate the merger rates, then these rates become largely consistent. The remaining differences between the merger rates are explained by the differences in the ranges of mass ratio measured by different techniques and the differing parent galaxy selection. 

Additionally, for mass-limited samples ($M_*>10^{10} ~\rm M_{\odot}$), they found excellent agreement between their merger rates from close pairs to several theoretical merger rates. Specifically, they agreed with the merger rates of \cite{hopkins10}, who, using a combination of semi-empirical models and high-resolution merger simulations, concluded that there are enough major mergers, to within a factor of $\sim2$, to account for the observed growth of the bulge population. They argue that previous studies reached different conclusions because they assumed incorrect merger timescales, rather, if a uniform simulated-calibrated merger timescale is used, then many of these works actually come to their conclusion \citep[see also][]{robaina10}. 

An alternative way to address whether there are enough mergers is to examine the phase in galaxy evolution that is predicted to correspond to the period of mergers. Adopting a simplified model in which blue star-forming galaxies merge and transform into red quiescent galaxies, one would expect a short period in which galaxies have intermediate colors, i.e., they are in the green valley. A recent study on the morphologies of green valley galaxies by \cite{mendez11} found that only 14\% of their sample are identified as on-going major or minor mergers (using G/$M_{20}$ and asymmetry parameters), which is lower than the 19\% merger rate in the blue cloud.  They further found that most green valley galaxies have disks and that 21\% have $B/T<0.05$, implying that they were unlikely to have experienced a recent major merger.

To conclude, while the merger model fits many aspects of our data, there are serious open questions, including whether major mergers are able to quench, whether there are enough of them, and whether they are consistent with the color and morphologies of green valley galaxies.  In the remaining discussion, we explore other models that may alleviate these problems.    

\subsection{Disk Instabilities: Violent and Otherwise}

An alternative bulge-building process involves the growth of giant clumps formed via gravitational instability in gas-rich disks \citep{noguchi99, elmegreen08, dekel09, ceverino10}. These clumps migrate inward and coalesce to form a bulge, and simulations suggest that galaxies can develop classical bulges with $n\approx4$ during this process\footnote{However, a recent paper by \cite{inoue11} claims that these clump-origin bulges are more akin to pseudobulges in that they exhibit $n<2$.} \citep{elmegreen08}. Recent simulations also show that the these clump-origin bulges have central surface mass densities comparable to that of our red sequence galaxies (Ceverino et al., in prep). The effectiveness of this instability, however, is highly dependent on the gas inflow rate onto the galaxy \citep{dekel09}, which declines with time. Thus, we expect this process to be more important at redshifts higher ($z \sim 2$) than that of our sample.  
 
Although this process may have operated strongly at $z\sim2$, we stress that this paper concerns a different sample of galaxies at lower redshift when conditions may have changed.  \cite{bell04} and \cite{faber07} found that the number of red sequence galaxies has at {\it least} doubled from $z\sim1$ to $z\sim0$, indicating that a fraction of our galaxies at $z\sim0.65$ are actively migrating to the red sequence as we view them. Using the NEWFIRM survey, \cite{brammer11} found that the mass density of quiescent galaxies with $M_*\gs3\times10^{10}~{\rm M_{\odot}}$ increased by a factor of $\sim10$ from $z\sim2$ to the present day. Similarly, \cite{hopkins10b} argue that the vast majority of ellipticals/spheroids do {\it not} form through high redshift channels. They state that the observed mass density of bulge-dominated galaxies at $z\sim2$ is only $\sim5\%$ of its $z=0$ value, and at $z\sim1$, is still only $\sim20-35\%$ of its $z=0$ value. Thus, most bulges are formed at $z \ls 1$, meaning the majority of our red sequence galaxies are recent arrivals. 

Although we expect violent disk instabilities to be increasingly less frequent at decreasing redshift, owing to lower gas fraction, the actual bulge contribution due to this mechanism at $z < 1$ is presently unknown. Hints at intermediate redshift suggest that the process may not be limited to high $z$.  For example, \cite{bournaud11} found that {\it half} of a sample of $z\sim0.7$ disk galaxies are clumpy without any merger signatures, implying that disk instabilities could still be important at that redshift.  

Perhaps we need to think more broadly and to recognize that the settling of matter to form regular, axisymmetric disks is a very lengthy process lasting many billions of years.  When any non-axisymmetric forces exist, an inevitable outcome is that {\it some} mass will be driven to the center.  Furthermore, in a general way the higher the degree of non-axisymmetry in the potential, the higher the rate of matter inflow will be.  At late times, non-axisymmetry has become small and the flow rate is low, a process that we call ``secular evolution'', which we discuss next. 

Future studies will resolve this question.

\subsection{Secular Evolution}

Secular evolution \citep{kk04} constitutes the weak end of the spectrum of non-axisymmetric processes in disk evolution.  This complex of processes involves the slow rearrangement of gas (and stellar) mass due to gravitational interactions between the gas and stars within a disk galaxy. A variety of relatively weak non-axisymmetric disturbances, such as bars, ovals, and spiral structure in the stellar component, can create non-central gravitational forces that add or subtract angular momentum from the gas, which responds by moving inward or outward depending on radius. The process sweeps inner gas into the center, where it forms stars, and pushes gas farther out to larger radii, where it can accumulate in tightly wrapped spiral arms or a ring \citep{simkin80,kk04}.  Separately, the gas itself may become mildly gravitationally unstable, which raises the velocity dispersion and causes the gas to radiate. This net loss of energy must come from somewhere, and the gas responds by sinking slowly to the center, increasing its (negative) potential energy \citep{forbes11}.  Overall, these processes push some mass to the center and other mass to the outskirts, thus increasing $n$. The forces are, however, weak and the process is slow, hence the term ``secular evolution''. The total time required would be many dynamical timescales, and thus at least several Gyr \citep{kk04, fisher09}.

Although secular evolution may contribute to some of the bulge-building taking place at $z \sim 0.65$, we do not believe it is the major factor. According to \cite{kk04}, bars are a major driver of secular evolution at the current epoch. Assuming that bars are also the main drivers at higher redshifts, comparing the bar fraction from the past, which is 10\%-25\% among late-types at $z > 0.8$  (\citealt{jogee04, sheth08}, Herrington et al., in prep), to the current epoch, which is 30-60\% at $z \sim 0$ \citep{sheth08, cameron10, masters11}, implies that secular evolution was not a major bulge-building process at $z\sim0.65$. Additionally, \cite{koo05} showed that $85\%$ of luminous field bulges within this redshift range are red, arguing against secular evolution being the dominant bulge-building process since they are expected to mainly produce blue bulges \citep{kk04}. 

Finally, the physical structure of bulges built by secular evolution differs strongly from ones built by mergers, as reviewed by \cite{kk04}.  The so-called ``classical'' bulges built by mergers resemble small ellipticals embedded in disk galaxies.  They have high stellar velocity dispersion and high vertical extent above the plane, having been ``fluffed up'' by the merger -- in other words,  they are true spheroids.  They can also be very massive and contain a considerable fraction of the total mass of the galaxy.  In contrast, the ``bulges'' built by secular evolution are relatively flat, have effective radii of only a few hundred pc, and have low fractional masses.  Because of these differences, \cite{kk04} term these structures ``pseudobulges''. \cite{drory07} directly compare the properties of these two types of bulges. They isolated a sample of nearby, massive, disk galaxies and classified them into classical bulges and pseudobulges based on the morphology of the central regions. Confirming their disparate nature, they found a clear bimodality in that pseudobulge galaxies are much bluer (in the blue cloud or green valley), have low central surface brightness, and have low global S\'ersic index ($n < 2.5$). 

For these reasons, we conclude that the pseudobulges cannot be an important contributor to our intermediate redshift, high $n$, red sequence galaxies. They may, however, certainly help build the bulges seen in late-type galaxies, becoming increasingly more important with decreasing redshifts, where the bar fraction can be as high as $60\%$ \citep{sheth08}\footnote{This number is controversial, recent studies by \cite{nair10, masters11, lee11} show that the local bar fraction is $\sim30\%$.}. In fact, \cite{fisher11} show that, by number, $80\%$ of the bulges within 11 Mpc of the Milky Way are actually pseudobulges. But, by mass, they only make up $\ls 10\%$ of the total mass density in local spheroids \citep{allen06, driver07}.

In conclusion, we find it helpful to think of the entire family of bulge-building mechanisms as ordered along a ``disturbance continuum'' from severe to mild, with corresponding timescales from short to long and bulge-building rates from fast to slow.  The members of this continuum consist of major to minor mergers at the strong end, through violent disk instabilities, to milder disk instabilities, and finally to weak, secular instabilities like bars, spiral arms, and normal star formation. The dividing line between an externally triggered process like mergers and internally triggered dynamical instabilities is fairly clear, but there is no such dividing line among the internal processes -- each one shades smoothly into the next.  However, since every disk is sooner or later subject to one or more of these processes, the central densities of disks {\it inevitably} tend to increase -- the only question is, how fast?  

We note that this disturbance continuum is also a continuum in time, with early galaxies experiencing disturbances at the strong end of the spectrum and later galaxies settling down to slower, more secular rates.  Our AEGIS galaxies exist somewhere near the middle of the time continuum, when disky galaxies as a class were considerably more disorganized and more non-axisymmetric than they are today.  This logic further supports our conclusion above that secular processes were probably not the major bulge-building process in these galaxies, with a combination of mergers and stronger disk instabilities being more likely.  However, the exact balance of these two processes remains for further study.

\subsection{Morphological Quenching}

The preceding sections focused on mechanisms to increase the central densities of disk galaxies, and thus account for one of our major findings, namely, higher central stellar densities in quenched galaxies.  Even though a variety of stellar build-up mechanisms were identified, including mergers and internal instabilities, we tacitly assumed regardless of process that high density would always lead to the creation of a black hole and that feedback from the black hole would quench star formation.  However, the discussion in \S 5.1 noted a lack of conclusive evidence that AGN feedback actually quenches star-formation in disk galaxies.  In this section and the next, we review two other quenching mechanisms that have been proposed to operate in central galaxies.

The first of these is morphological quenching \citep{martig09}, whereby the steep potential well of a bulge is able to drive the Toomre Q parameter above unity and stabilize the gas disk against star formation.  An attractive aspect of morphological quenching is that it sets in when $n$ is high, which jibes with the structure of quenched galaxies. This mechanism is unique because it does not require the removal of gas or the suppression of the cold-gas supply onto the galaxy.  Instead, gas can continue to accrete onto a galaxy yet remain inert to star formation owing to the strong central mass concentration. A recent analysis of a set of three high-resolution AMR simulations at $z \simeq 2.3$ by \cite{ceverino10} demonstrates this process. From $z\simeq 2.3$ to $z\sim1$, two of these galaxies are shown to evolve from a gravitationally unstable and turbulent disk into a stable system; they attribute this change to presence of a dominant stellar bulge. 

An explicit prediction of morphological quenching is that red, early-type galaxies could host significant cold gas in the amount of a few percent of their baryonic mass, and in fact comparable to gas fractions in normal star-forming galaxies \citep{martig09}.  However, this prediction is not consistent with observed H I properties of nearby early-type galaxies.  Though H I is frequently detected, especially in field galaxies \citep{morganti06, oosterloo10}, the amounts are nearly always low.  This is confirmed by the GASS H I survey at Arecibo of $\sim1000$ slightly more distant massive galaxies with $M_* > 10^{10} ~ \rm M_{\odot}$ \citep{catinella10}, which shows that the average H I fractions of red sequence galaxies are at least ten times lower than galaxies of similar mass on the star-forming main sequence \citep{schiminovich10, fabello11}. The same result applies to molecular H$_2$ in the same galaxies \citep{saintonge11}. These recent studies simply confirm what has been known for a long time about early-type galaxies, that their absolute gas contents are low compared to star-forming galaxies \citep[e.g.,][]{vandriel91, roberts94, noordermeer05}. Hence, in order to achieve quenching, it is necessary to reduce the fractional gas content, either by expelling gas or preventing new gas from falling in -- morphological quenching alone cannot do the job.

For the nearest galaxies it is possible to map the H I distributions \citep{vandriel91, noordermeer05, morganti06, oosterloo10}.  Normal lenticular galaxies typically have H I in an outer ring with an empty hole in the middle.  Several authors \citep{vandriel91, cortese09, oosterloo10} have stressed that the surface density of gas in these rings is well below the critical threshold for star formation \citep{kennicutt89, schaye04, bigiel08}, which represents the threshold for molecular H$_2$ formation, a tracer of gravitationally unstable gas  \citep[e.g.][]{krumholz11}.   Moreover, the predicted star formation efficiency at such low densities is ten times lower than in normal star-forming galaxies \citep{krumholz12},  which agrees well with the low star-formation efficiencies seen in the Arecibo GASS survey \citep{schiminovich10, fabello11}.  Such low-level star formation has recently been detected in the outer disks of normal green-valley S0 galaxies in {\it HST} U imaging (\citealt{salim10, salim12}; Fang et al., in prep). 

Hence, we reach a very important conclusion about the evolutionary track of quenching: low gas content is the underlying cause of quenching, but star formation shuts down even faster as gas content falls below the threshold value, owing to the nonlinear relation between cold gas surface density and star formation rate. This fall in star-formation efficiency causes galaxies to redden even faster than expected, propelling them rapidly to the red sequence\footnote{An interesting corollary comes from the fact that UV colors are more sensitive to weak star formation than optical colors, and thus galaxies can be on the red sequence according to $U-B$ but in the green valley according to $NUV-r$; this is actually seen (e.g., \citealp{salim10, salim12}).  Since galaxies in this paper are classified using $U-B$, it is possible that some of our red sequence objects would appear in the green valley if near-UV color were used.}. 

It is fair to point out that all of the data cited above to evaluate morphological quenching comes from nearby galaxies, some of which are members of virialized clusters and dense groups that are subject to environmental processes such as ram-pressure stripping or strangulation.  However, many of the nearest galaxies are known {\it not} to be in clusters \citep[e.g.,][]{oosterloo10}, and there must also be many field objects amongst the thousand or so galaxies in GASS, yet the trends are the same.  In short, we cannot think of any reason why morphological quenching would be the key causative agent for quenching at higher redshift when it does not appear to play that role (even for field galaxies) today.

\subsection{Critical Halo Mass} \label{sub:halomass}

The termination of the cold gas supply due to a critical halo mass is commonly referred to as `halo quenching' \citep{silk77, rees&ostriker77, blumenthal84, birnboim03, keres05, dekel06, cattaneo06}. Halos below $\sim 10^{12} ~\rm{M_{\odot}}$ are able to accrete gas through cold flows while halos above the threshold mass experience a virial shock that heats the gas. The hot, diluted gas in massive halos is vulnerable to feedback from AGN, which effectively halts star formation. An interesting outcome of this theory is the ability to generate a hot halo atmosphere that allows for AGN `radio mode' feedback \citep{croton06, dekel06}. This mechanism provides a way to permanently quench a galaxy, which is desirable since the stellar populations of most local ellipticals show no signs of recent star formation (since at least $z\sim1$; \citealt{thomas05}), despite continual gas infusion by stars.

Various studies have explored this theory. For example, using a data-driven, halo-abundance matching technique that spans $0<z<1$, \cite{conroy09} found a gradual transition of galaxy properties -- including specific star formation rate -- across a halo mass of $\sim 10^{12}~\rm{M_{\odot}}$. Recently, \cite{more11} inferred the halo mass-stellar mass relation based on kinematics of SDSS satellite galaxies and found that red, central galaxies, on average, occupy more massive halos than blue centrals for fixed luminosity, but show a less appreciable difference for fixed stellar mass. \cite{woo12} study the quenched fraction of central galaxies as a function of halo mass and find that the span of halo masses between 20\% quenched fraction and 60\% quenched fraction is a whole 1.5 dex. This agrees well with the width of 0.8 dex seen here in the overlap region using stellar mass (and the same quenched fractions) in Fig.~\ref{fig:findcrit_all}b and the theoretical scaling law between central and halo mass \citep{kang05, cattaneo06}.  By contrast, the width using our preferred parameter, inner surface density, is only 0.5 dex (cf. Fig. 5f).  An interpretation that emerges from these works is that halo mass has a gradual and probabilistic effect upon galaxy properties, which is therefore very consistent with a wide overlap region in halo mass. Instead of a sharp transition at $\sim 10^{12} ~\rm{M_{\odot}}$, galaxies seem to become ready for star-formation quenching around this critical value, followed by some type of event that ultimately triggers quenching.

This scenario ties in well with our two-stage picture in which galaxies ``ripen'' along the blue cloud, becoming more and more susceptible to quenching as they age. In \S\ref{sub:mergers}, we associated this ripening with decreasing gas content with stellar mass in the blue cloud, which is also seen in our data as an increase in $U-B$ with stellar mass (Fig.~\ref{fig:findcrit_all}b). Lower gas content means that less gas needs to be removed in quenching, which means in turn that more massive blue cloud galaxies are more vulnerable to quenching. Their lower gas content stems from the fact that their halo masses are closer to the critical value $\sim 10^{12} ~ \rm M_{\odot}$, the neighborhood where cold accretion shuts down.  This basic picture does not change if $M_*/r_{\rm e}$ \citep{franx08} or central velocity dispersion \citep{wake12b} is substituted for stellar mass -- any parameter that tracks halo mass reasonably well can serve as a ripeness indicator.  

This logic leads to a picture in which the changes in gas fraction, color, and star formation rate along the blue cloud are caused by the gradual dominance of shock-heated gas over cold streams as galaxies near the critical halo mass. A tougher challenge, though, is to explain why the actual quenching state relates so closely to conditions {\it at the center of the galaxy} -- why is this link so close if the primary governor of cooling is out in the halo?

We have no firm answer to this but offer some speculations.  Evidently the central conditions either signal, or even trigger, a second quenching mechanism and this, plus ``natural'' halo quenching, is what tips a galaxy over the edge.  The obvious candidate for this second process is AGN feedback, but we have stressed that direct evidence for this is still weak.  It is good to be cautious since the ERIS Milky Way simulation \citep{guedes11} develops a red bulge and high central stellar density quite naturally through mergers and/or internal disk evolution.  Its star formation rate is falling and it seems well on its way to the red sequence, all without benefit of AGN feedback. 

Ideal would be some mechanism that both correlates with central density and can switch a halo quickly from cold mode to hot mode.  Some possibilities come to mind.  Perhaps AGN feedback helps to heat the halo.  Perhaps exhaustion of gas at the center enhances the ability of stellar winds to sweep gas out of the galaxy.  Finally, perhaps a merger simultaneously builds up central stellar density {\it and} triggers a full standing shock.  Such a transformation is seen in simulations (Dekel et al., in prep) where a minor merger triggers a shock that expands from the halo center to the virial radius and heats the medium.  The larger point is that quenching is probably not just one factor but a combination of factors that build to some critical threshold.

\subsection{The Relationship Between Color and Star Formation} \label{sub:issues}

In this section, we broach the lingering issue of the relationship between color and star formation. Throughout this paper, we have constantly interchanged these two parameters, suggesting that color is a good representation of specific star formation rate. However, the issue of dust has not been thoroughly addressed in our colors. Although we have excluded edge-on galaxies (that are presumably highly affected by dust) and ensured that our red sequence galaxies are truly quiescent (using the $UVJ$ diagram; see \S\ref{sec:selection}), we have not actually made any dust correction to our rest-frame $U-B$ colors. Therefore, the color trends that we have examined throughout this paper may not exactly translate into star formation trends. The analysis most affected would be our interpretation of how the star formation rates of galaxies behave within the blue cloud. We remarked on the tight trend between color and $M_*/r_{\rm e}$ (and $M_*$) within the blue cloud in Figs.~\ref{fig:findcrit_all}b (and \ref{fig:findcrit_all}c) and we proposed that star formation must decrease with $M_*/r_{\rm e}$ (or $M_*$). However, this color trend may instead be due to more dust in larger galaxies. If this is the case, then a galaxy's star formation would be independent of $M_*/r_{\rm e}$ (or $M_*$) within the blue cloud. However, work using dust-corrected or dust-robust multi-wavelength data \citep[e.g.,][]{noeske07, salim07, schiminovich07, zheng07} have shown that specific star formation rate does indeed decrease with increasing stellar mass, which is what we have inferred from our color-mass diagram (Fig.~\ref{fig:findcrit_all}b). Thus, these works justify our subtle assertion that color is a proxy for star formation. Most importantly though, our lack of dust-corrected $U-B$ colors do not affect our main result that the inner stellar structure of galaxies is most related to quiescence, since our quenching analysis is based on differentiating galaxies on the blue cloud from those on the red sequence, which we have ensured to be unaffected by dust (see \S\ref{sec:selection}).

\section{Conclusion}

In this paper, we analyze a sample of DEEP2/AEGIS galaxies in the redshift range $0.5 \le z < 0.8$ using $HST/$ACS $V,I$ images. Our sample has been run through GIM2D, a bulge+disk decomposition program that gives us information on the subcomponents of intermediate redshift galaxies. Using these data, which we provide at \url{http://people.ucsc.edu/~echeung1/data.html}, our goal is to address how quenching depends on galactic structure.

Our methodology is to assess the color correlations of several structural parameters -- $M_*$, $M_*/r_{\rm e}$, $M_*/r_{\rm e}^2$, $n$, and $\Sigma_{\rm 1 kpc}^*$ -- by computing an `overlap region', which is the band in color-parameter space that encompasses both red and blue galaxies. To quantify overlap regions, we calculate the fractional number of galaxies within the overlap region $\Delta$N/N and the fractional extent of the region $\Delta$x/x; the parameter with the smallest overlap region is considered to be the best predictor of quiescence. Finding that no parameter is a perfect predictor of quenching, we explore the overlap region of $M_*/r_{\rm e}$ for secondary color correlations amongst a variety of other parameters, including those of bulge and disk. We also consider the number of severe outliers from the best predictor of quiescence, $n$. Our results are the following:

\begin{enumerate}

 \item The S\'ersic index ($n$) most sharply discriminates between the red sequence and the blue cloud. Eliminating dusty, red sequence contaminants ensures that $n$ targets quiescent galaxies, not merely red ones. Moreover, the color-$n$ diagram resembles a step function, suggesting that $n$ is related to a physical quenching threshold.

 \item However, there are exceptions to this general behavior -- blue galaxies make up $\approx40\%$ of our $n>2.3$ galaxies. Suspecting contamination from starbursts, AGNs, and/or other sources of error, we measured central surface stellar mass densities, which revealed that these outliers do not truly belong with the red sequence -- their $\Sigma_{\rm 1 kpc}^*$ are much lower. Central surface density corrects these outliers and hints that it is the {\it inner} structure of galaxies that is most related to quiescence.
 
 \item Red sequence bulges are roughly twice as massive as blue cloud bulges at the same galaxy stellar mass (and $M_*/r_{\rm e}$), yet also roughly twice as small.  This structural difference shows that most blue galaxies at the observed epoch do not simply fade onto the red sequence.  Rather, the high values of $n$ and $\Sigma_{\rm 1 kpc}^*$ on the red sequence must reflect a net migration of existing stars toward the center of the system or the formation of new stars at the center.  This restructuring either causes quenching itself or is closely related to the process that does. 

 \item While in the blue cloud before quenching, a galaxy's star formation rate is most closely correlated with $M_*/r_{\rm e}$ (or with $M_*$).  

\end{enumerate}

These results suggest that galaxies evolve toward the red sequence in a two-stage process. In stage one, a galaxy is star-forming in the blue cloud at a rate that correlates with global parameters like $M_*/r_{\rm e}$ (or $M_*$). Since these in turn reflect halo mass, a broad conclusion might be that star-formation is controlled by the galaxy's halo while stars are still forming.  As stellar mass increases, the halo mass also increases, ultimately approaching the critical value $\sim 10^{12}~ \rm M_{\odot}$, where cold flows begin to give way to hot gas, which cannot accrete.  The gas content of the galaxy begins to fall as it nears this critical value, and colors redden.  

The result of stage one is a galaxy that is increasingly vulnerable as time goes by to the onset of a second quenching process.  This second process must be closely associated with bulge-building, and central stellar density, $\Sigma_{\rm 1 kpc}^*$, must increase.  AGN feedback is an obvious candidate for this second process, but direct evidence for this is still weak.  Also unclear is the exact mechanism that leads to the central mass build-up at the center of the galaxy.  We have stressed that galaxies, particularly at high redshift, are far from axially symmetric and that any non-axisymmetry leads inevitably to an exchange of angular momentum and/or loss of energy, which causes some stars and gas to move inward.  Major mergers sit at the strong end of this ``disturbance continuum'', secular evolution processes sit at the other end, and minor mergers, violent disk instabilities, and milder disk instabilities sit in the middle. Given that our galaxies lie at $z \sim 0.65$, where galaxies are still moderately disturbed, it is unlikely that secular evolution plays a major role in them. More likely is some combination of mergers and disk instabilities, which collectively are probably strong enough and frequent enough to do the job.  Mergers in particular have several well known advantages: they build bulges at the right stellar mass, they naturally build a concentrated stellar spheroid, they drive a lot of gas quickly to the center that can power an AGN or starburst, and they may be able to quickly switch a marginally cooling halo into hot mode. A problem for major mergers is the large fraction of galaxies in the green valley that are disky, but these might be explained by the other mechanisms or they might be reddened members of the blue cloud.  

One conclusion seems clear, and that is that moving into the green valley and thence to the red sequence requires a lowering of fractional cold gas content. This can only be achieved either by expelling gas or by preventing its infall.  Exactly how this happens is still not clear, but at least some of the parameters surrounding the process are better known.       

\acknowledgments

We thank Thiago Signorini Gon\c{c}alves, Stijn Wuyts, Joanna Woo, Alexis Finoguenov, Jonathan Trump, and Lia Athanassoula for useful comments. We would also like to thank the anonymous referee for a very helpful report that resulted in substantial improvements to the paper.

The DEEP2 survey was initiated under the auspices of the NSF Center
for Particle Astrophysics. Major grant support was provided by
National Science Foundation grants AST 95-29098, 00-711098, 05-07483,
and 08-08133 to UCSC and AST 00-71048, 05-07428, and 08-07630 to UCB.
The DEEP2 survey has been made possible through the dedicated efforts
of the DEIMOS instrument team at UC Santa Cruz and support of the
staff at Keck Observatory.  The {\it HST} ACS mosaic in EGS was
constructed by Anton Koekemoer and Jennifer Lotz and was funded by
grant HST-AR-01947 from NASA.  Finally, we recognize and acknowledge
the highly significant cultural role and reverence that the summit of
Mauna Kea has always had within the indigenous Hawaiian community; it
has been a privilege to be given the opportunity to conduct
observations from this mountain.

\appendix

\section{Surface Brightness Limits}
\label{app:sb}

\begin{figure*}[t!] 
\centering
\includegraphics[scale=.7]{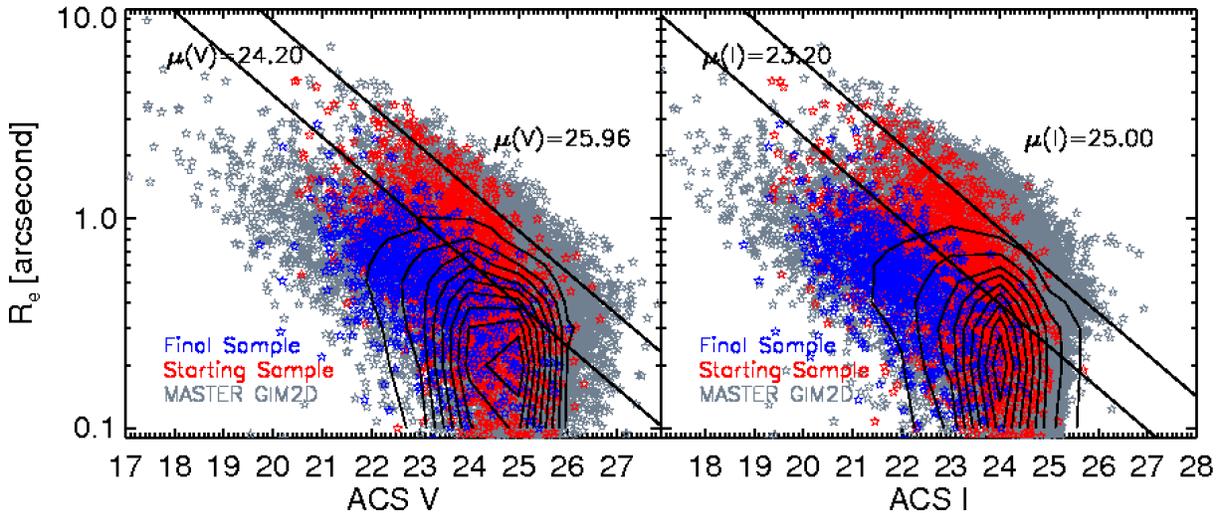}
\caption{Semimajor axis effective radius in arcsecond is plotted against $V$ and $I$. The two lines of constant surface brightness in each panel roughly marks the edge of the `starting' and `final' samples' surface brightness distribution. The contours represent the number density of the master GIM2D sample. 
\label{fig:surf_bright}}
\end{figure*}

One might be concerned that our `final' sample may be missing low surface brightness galaxies; in this Appendix, we show that this is not the case. We demonstrate our `final' and `starting' samples' surface brightness limits in Fig.~\ref{fig:surf_bright} by plotting the effective radius in arcseconds vs. $V$ and $I$ (see \S\ref{sec:selection} for sample definitions). The master GIM2D sample is plotted in gray and the `starting' and `final' sample are overplotted in red and blue, respectively. In each panel, we draw two lines of constant surface brightness to give an idea of our samples' surface brightness boundaries. There is obviously an edge to the distribution of the master GIM2D sample at low surface brightnesses. Whether this is a selection limit or a reflection of where real galaxies lie is unclear. But it is clear that the master GIM2D sample contains galaxies with the lowest surface brightnesses, indicating that our `starting' and `final' samples are sensitive to them.  

In these plots, the relevant difference between the `starting' and `final' sample is the imposed mass limit of the `final' sample. The main concern in terms of sample completeness boils down to whether the `final' sample is missing massive low surface brightness galaxies. It is clear that there are many galaxies in `starting' sample (and the master GIM2D sample) with much lower surface brightnesses than those in the `final' sample. Hence if such massive low surface brightness galaxies existed, the `final' sample should certainly contain them, thus our `final' sample does not miss low surface brightness galaxies.

\section{GIM2D Measurement Quaility}
\label{app:gim2derror}

\begin{figure*}[t!] 
\centering
\includegraphics[scale=.7]{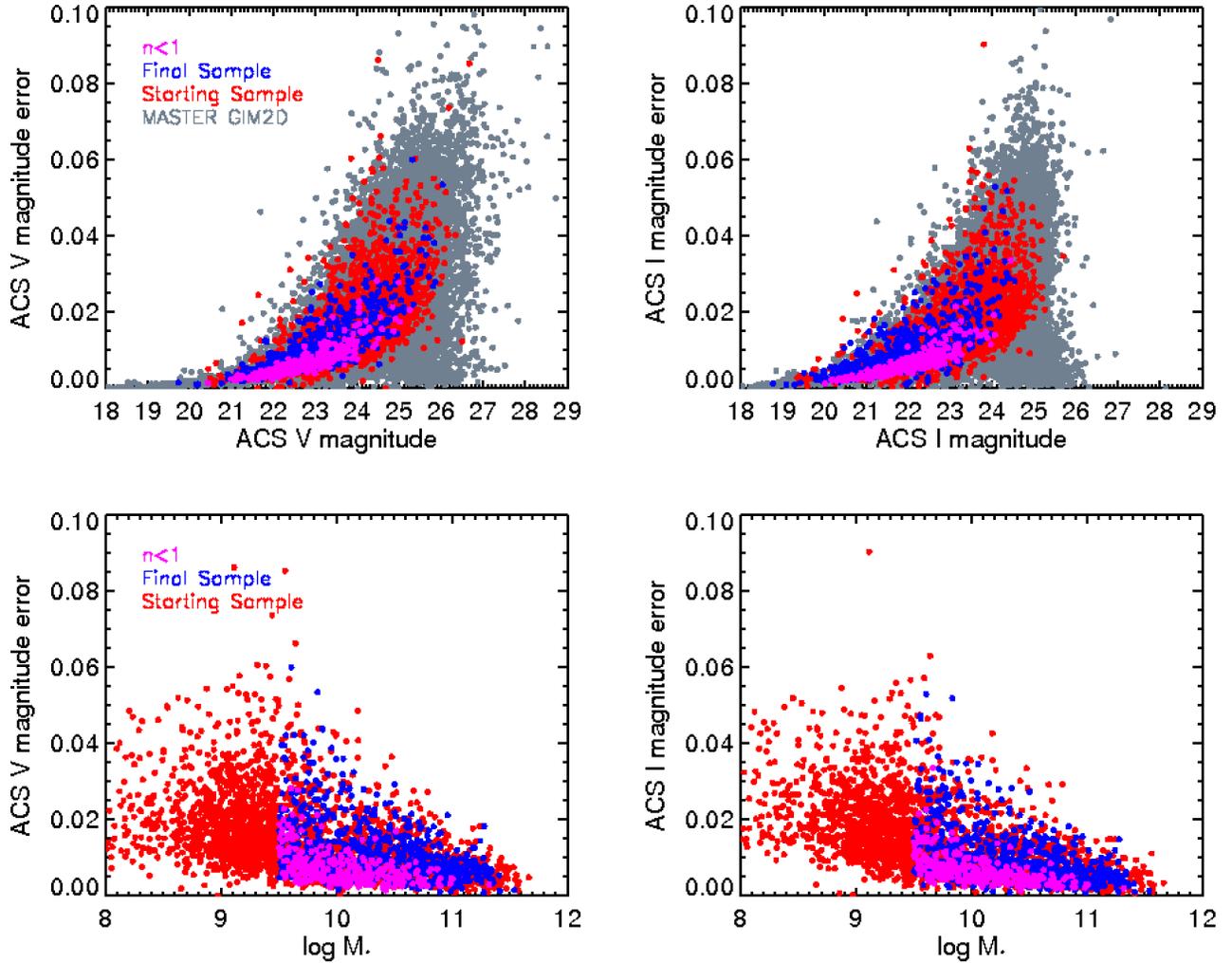}
\caption{$V$ and $I$ magnitude errors as a function of $V,I$, and $\log~M_*$.
\label{fig:mag_errors}}
\end{figure*}

In this appendix, we show that the errors on the GIM2D measurements are well-behaved even at the sample limits. The most important quantities are the GIM2D model $HST$/ACS $V$ and $I$ magnitudes since every structural measurement is based on them. Fig.~\ref{fig:mag_errors} plots the $V$ and $I$ 1-$\sigma$ errors as a function of $V$, $I$, and $\log~M_*$. These errors are confidence limits derived through full Monte Carlo propagations of the parameter probability distributions computed by GIM2D \citep{simard02}. 
 
Shown are the master GIM2D catalog (gray), `starting' sample (red), `final' sample (blue), and galaxies within the `final' sample with S\'ersic indices $n<1$ (pink). The 5-$\sigma$ limiting magnitudes for the $HST$/ACS images are $V=26.23$ and $I=25.61$, consistent with the master GIM2D distribution. Within our `final' sample, the limits of the magnitude distributions are $\approx0.5$ mag and $\approx1.0$ mag brighter than those limits, respectively. And most importantly, the $V$ and $I$ magnitude errors at the mass and magnitude limit of the `final' sample are small, $ \le 0.06$ mag, with the majority $\le 0.03$ mag.
 
\begin{figure*}[t!] 
\centering
\includegraphics[scale=.7]{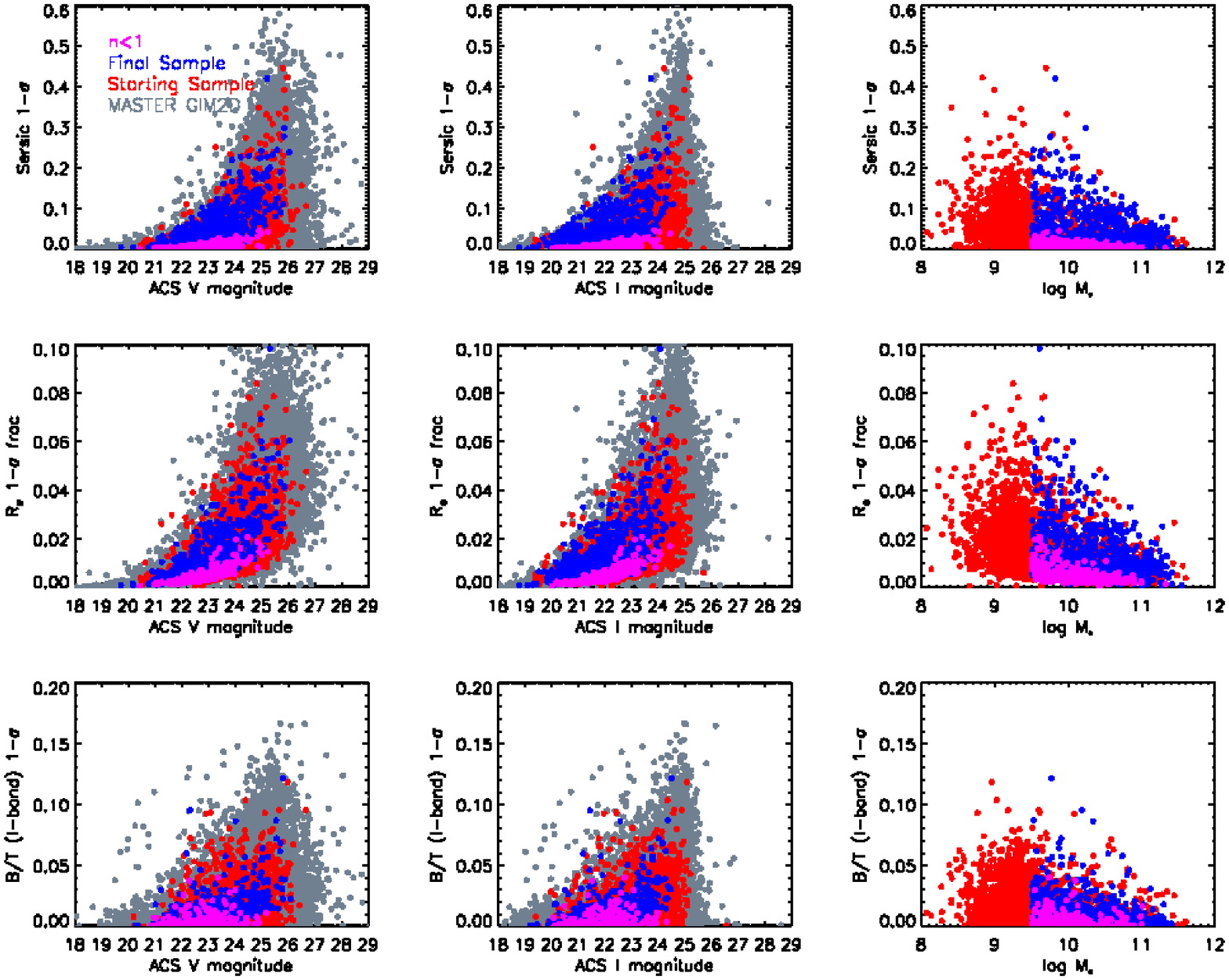}
\caption{S\'ersic index error, effective radius fractional error, and bulge-to-total ratio errors as a function of $V,I$, and $\log~M_*$.
\label{fig:sersic_errors}}
\end{figure*}

To further illustrate the quality of the `final' sample's GIM2D fits, we plot in Fig.~\ref{fig:sersic_errors} the 1-$\sigma$ error of the S\'ersic index, fractional 1-$\sigma$ error of the effective radius, and the 1-$\sigma$ error of the bulge-to-total ratio $B/T$ (in the $I$ band) as a function of $V$, $I$, and $M_*$. The errors of the S\'ersic index measurements are almost all $\le 0.2$. A typical late-type galaxy has $n\sim1$ while a typical early-type galaxy has $n\sim4$. An error of $0.2$ on $n$ will not affect this division and hence these errors are tolerable. The errors on the effective radius measurements are presented as fractional errors, i.e., 1-$\sigma$ effective radius error divided by the effective radius. Note that these radii are semimajor axis effective radii and not circularized effective radii. The fractional errors are small, almost all are $\le 6\%$. The errors on the $B/T$ are almost all $\le 0.05$. Most star-forming, blue galaxies in our `starting' sample, which are presumably late-type, have $B/T \le 0.10$ while most quiescent, red galaxies in our `starting' sample, which are presumably early-type, have $B/T \ge 0.40$. Thus an error of $0.05$ will not push blue galaxies to $B/T$ values of red galaxies and vice versa. In summary, these plots demonstrate that our `final' sample contain quality GIM2D measurements down to the mass and magnitude limits of the `final' sample.

 \section{S\'ersic Index Bias?}
 \label{app:sersicindex}

\begin{figure*}[t!] 
\centering
\includegraphics[scale=.7]{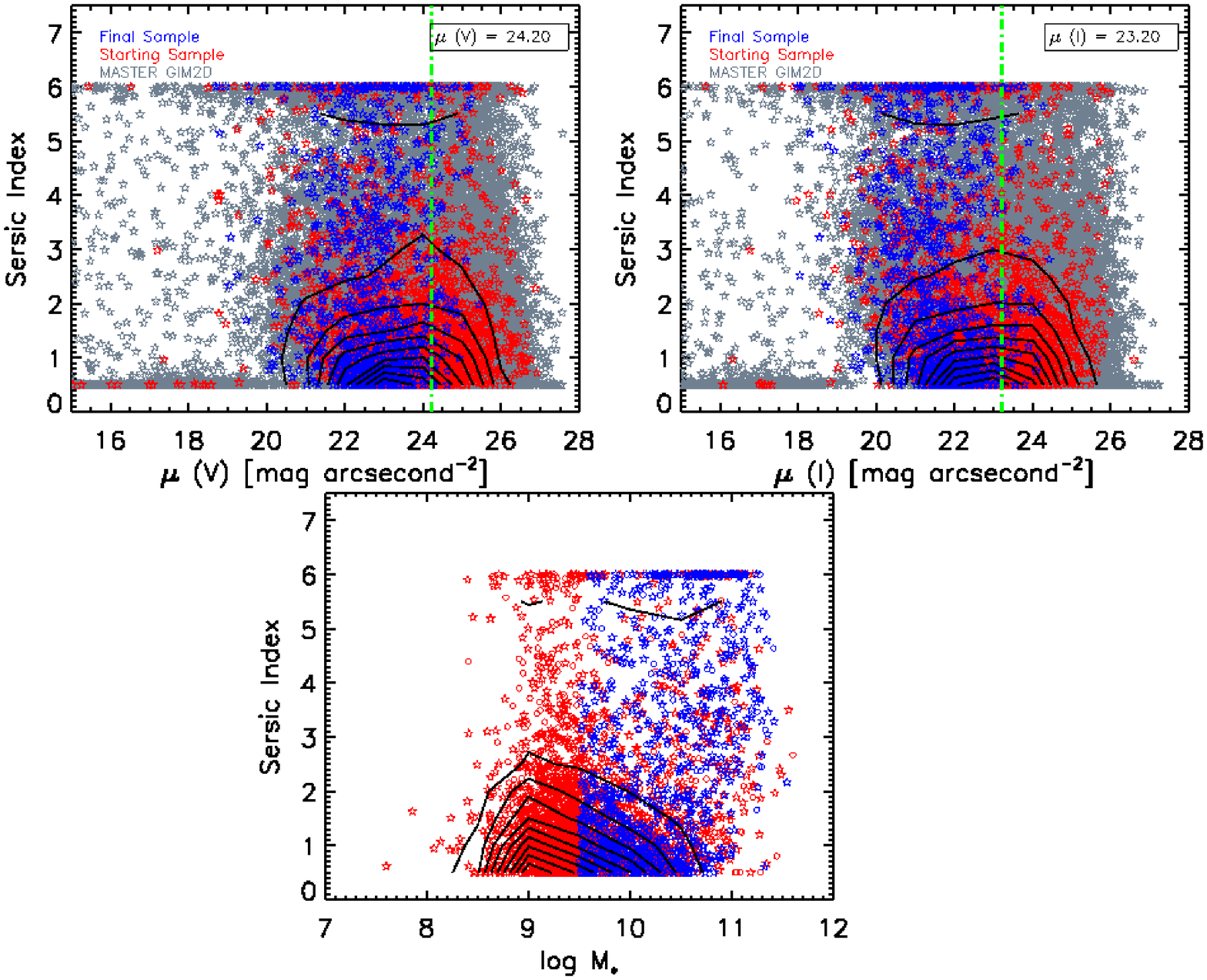}
\caption{S\'ersic index is plotted against $\mu (V)$ and $\mu(I)$ in the top row. Contours represent the number density of the GIM2D master sample in gray. The vertical green dot-dashed line represents the approximate edge of our `final' sample's surface brightnesses distribution (as seen in  Fig.~\ref{fig:surf_bright}). The bottom panel plots the S\'ersic index against stellar mass. The `final' sample does not show a strong disparity of high and low $n$ populations at faint (low) surface brightness (mass).
\label{fig:surf_brightvsseric}}
\end{figure*}

One might be concerned that the S\'ersic index may be sensitive to surface brightness and stellar mass. Fig.~\ref{fig:surf_brightvsseric} plots the S\'ersic index against these quantities. The top row is the $\mu$ in $V$ and $I$. The vertical green dot-dashed line represents the approximate edge of the `final' sample's surface brightness distribution as seen in Fig.~\ref{fig:surf_bright}. The contours represent the number density of the master GIM2D catalog, which clearly shows that the majority of these galaxies have low $n$. At faint $\mu$, the `starting' sample clearly has less $n>3$ galaxies compared to the $n<3$ population, but we would argue that is a reflection of the universe, i.e., high $n$ galaxies generally have bright $\mu$. For the `final' sample, however, this proportional dearth of $n>3$, faint $\mu$ galaxies seen in the `starting' sample seems to have been greatly reduced. 

We also plot the S\'ersic index against stellar mass in the lower panel. At low stellar masses, the `starting' sample also has less high $n$ galaxies than low $n$ galaxies. However, this absence does not seem as severe as with faint $\mu$. And within our `final' sample, it seems that this disparity in S\'ersic populations have decreased. Moreover, it is important to note that even at the lowest masses, our `final' sample still contains S\'ersic index values across the entire possible range, indicating that GIM2D does not preclude high $n$ fits for low mass galaxies. In fact, the absolute count of galaxies with S\'ersic index $n>3$ in every mass range is approximately the same. Hence there is no predisposition against $n>3$ galaxies in our `final' sample. The fact that our `final' sample has a large concentration of low $n$ galaxies is most likely due to the fact that there are many late-type galaxies at low masses.

One might wonder how the low $n$ galaxies fared near the limits of the data. As can be seen in Fig.~\ref{fig:surf_brightvsseric}, the low $n$ galaxies make up a significant fraction of our `final' sample. To see the reliability of the measurements of these galaxies, we've highlighted this population ($n<1$ galaxies in the `final' sample) in magenta in several of the previous figures, specifically, Fig.~\ref{fig:mag_errors} and \ref{fig:sersic_errors}. In most cases, these low $n$ galaxies have errors consistent with the rest of the population, i.e., low $n$ galaxies fare fine near the limits of our `final' sample. 

\bigskip



\newpage

\begin{deluxetable}{lccccccccc}
\tabletypesize{\scriptsize}
\tablewidth{0pt} 
\tablecaption{Galaxy Properties}
 \tablehead{   
\colhead{DEEPID} &
\colhead{$Q$}&
\colhead{$z$}&
\colhead{$M_{B} $} &
\colhead{$M_{B} $}&
\colhead{$U-B $} &
\colhead{$U-B $} &
\colhead{$M_*$} &
\colhead{$M_*$} &\\
\colhead{} &
\colhead{}&
\colhead{}&
\colhead{$k$-correct} &
\colhead{Eqns.~\ref{eqn:mb}, \ref{eqn:kcorrect}} &
\colhead{$k$-correct} &
\colhead{Eqn.~\ref{eqn:ub}}&
\colhead{J. Huang et al., in prep} &
\colhead{Eqns.~\ref{eqn:masslight}, \ref{eqn:mb}, \ref{eqn:kcorrect}} \\
\colhead{(1)} & \colhead{(2)} & \colhead {(3)} &
\colhead{(4)} & \colhead{(5)} & \colhead {(6)} &
\colhead{(7)} & 
\colhead{(8)} & \colhead{(9)}
}
\startdata      
         13049654 & 4 & 0.20 & -17.52 & -17.49 & 0.77 & 0.77 & 9.14 & 9.19 \\
    13018599 & 4 & 0.79 & -20.14 & -20.13 & 1.14 & 1.14 & 10.56 & 10.42 \\
    12020067 & 4 & 0.45 & -19.52 & -19.38 & 0.62 & 0.65 & 9.38 & 9.49 \\
    12007757 & 4 & 0.99 & -20.80 & -20.91 & 0.89 & 0.83 & 10.09 & 10.12 \\
    13040619 & 4 & 0.71 & -20.78 & -20.96 & 1.34 & 1.24 & 10.81 & 10.82 \\
    13048556 & 4 & 0.61 & -20.03 & -20.48 & 0.64 & 0.59 & 9.53 & 9.75 \\
    13049852 & 4 & 0.57 & -20.93 & -21.01 & 1.44 & 1.39 & 10.93 & 10.94 \\
    13026131 & -1 & 0.54 & -18.73 & -18.55 & 0.79 & 0.89 & 9.51 & 9.46 \\
    12020067 & 4 & 0.45 & -19.52 & -19.38 & 0.62 & 0.65 & 9.38 & 9.49 \\
    12015606 & 4 & 0.67 & -20.77 & -20.85 & 0.78 & 0.88 & 10.27 & 10.36 \\
    13058131 & 4 & 0.71 & -20.84 & -20.98 & 1.34 & 1.32 & 10.74 & 10.90 \\
    12016156 & 4 & 0.74 & -19.47 & -19.57 & 0.48 & 0.57 & -1.00 & 9.37 \\
    13012297 & 4 & 0.81 & -21.26 & -21.19 & 1.23 & 1.24 & 10.92 & 10.94 \\
    12007918 & 4 & 0.66 & -20.47 & -20.45 & 0.63 & 0.68 & 9.79 & 9.90 \\
    13011795 & -1 & 0.71 & -19.81 & -19.82 & 1.33 & 1.32 & 10.37 & 10.43 \\
    13064645 & 4 & 1.21 & -22.56 & -22.10 & 1.45 & 0.79 & 11.09 & 11.14 \\
    12016799 & 3 & 0.68 & -20.54 & -20.54 & 1.16 & 1.12 & 10.67 & 10.53 \\
    12023870 & -1 & 0.48 & -18.78 & -18.88 & 0.56 & 0.57 & 9.18 & 9.12 \\
    12004470 & -1 & 0.46 & -19.31 & -19.15 & 0.29 & 0.53 & 9.13 & 9.17 \\
    12020439 & 4 & 0.58 & -19.17 & -18.97 & 0.74 & 0.66 & 10.03 & 9.27 \\

    \enddata
\tablecomments{Twenty randomly-selected galaxies from the catalog is shown. The entire catalog is available at: \url{http://people.ucsc.edu/~echeung1/data.html}. 
Col. (1): Unique DEEPID. 
Col. (2): Spectroscopic redshift quality code. Values of 3 and above are secure spectroscopic redshifts (see \citealp{newman12} for more details) and they comprise the spectroscopic sample. For values below 3, we use the photometric sample and their corresponding photometric redshifts. 
Col. (3): Best available redshift  For those with $Q\ge3$, these are spectroscopic $z$, those with $Q\le2$ are photometric $z$.
Col. (4): Absolute $B$-band magnitude derived from $k$-correct. About $\approx7\%$ have large error measurements, we use vales from Col. (5) for these objects.
Col. (5): Absolute $B$-band magnitude derived from Eqns.~\ref{eqn:mb}, \ref{eqn:kcorrect}.
Col. (6): $U-B$ rest-frame color derived from $k$-correct. About $\approx7\%$ have large error measurements, we use vales from Col. (7) for these objects.
Col. (7): $U-B$ rest-frame color derived from Eqn.~\ref{eqn:ub}.
Col. (8): Stellar mass from J. Huang et al., in prep. $\approx 10\%$ of our sample have no corresponding $M_*$; they are marked by $-1.00$. We use Col. (9) for these objects.
Col. (9): Stellar mass derived from Eqns.~\ref{eqn:masslight}, \ref{eqn:mb}, \ref{eqn:kcorrect}.
}
\label{tab:1}
\end{deluxetable}

\begin{deluxetable}{lccccccc}
\tabletypesize{\scriptsize}
\tablewidth{0pt} 
\tablecaption{Subcomponent Properties}
 \tablehead{   
   \colhead{DEEPID} &
\colhead{$M_{B} $} &
\colhead{$M_{B} $} &
\colhead{$U-B $} &
\colhead{$U-B $} &
\colhead{$M_*$} &
\colhead{$M_*$} &
\\
\colhead{ }&
\colhead{Bulge} &
\colhead{Disk} &
\colhead{Bulge} &
\colhead{Disk} &
\colhead{Bulge} &
\colhead{Disk} &
\\
\colhead{(1)} & \colhead{(2)} & \colhead {(3)} &
\colhead{(4)} & \colhead{(5)} & \colhead {(6)} &
\colhead{(7)} 
 }
\startdata    
        13049654 & -15.95 & -17.22 & 1.15 & 0.73 & 8.99 & 9.02 \\
    13018599 & -19.17 & -19.64 & 1.53 & 1.05 & 10.27 & 10.12 \\
    12020067 & -18.85 & -18.68 & 0.73 & 0.61 & 9.39 & 9.14 \\
    12007757 & -20.30 & -20.55 & 1.00 & 0.66 & 10.18 & 9.63 \\
    13040619 & -19.62 & -19.88 & 1.07 & 1.52 & 10.11 & 10.57 \\
    13048556 & -19.15 & -19.52 & 0.54 & 0.66 & 9.13 & 9.49 \\
    13049852 & -19.33 & -20.34 & 1.34 & 1.42 & 10.23 & 10.70 \\
    13026131 & -17.65 & -18.11 & 1.24 & 0.74 & 9.48 & 9.07 \\
    12020067 & -18.85 & -18.68 & 0.73 & 0.61 & 9.39 & 9.14 \\
    12015606 & -19.84 & -20.20 & 1.20 & 0.60 & 10.33 & 9.66 \\
    13058131 & -19.60 & -20.08 & 1.50 & 1.06 & 10.45 & 10.29 \\
    12016156 & -18.85 & -18.91 & 0.67 & 0.41 & 9.25 & 8.79 \\
    13012297 & -21.05 & -19.37 & 1.25 & 1.18 & 10.89 & 10.16 \\
    12007918 & -19.91 & -19.75 & 1.01 & 0.50 & 10.15 & 9.29 \\
    13011795 & -18.66 & -18.93 & 1.37 & 1.18 & 10.00 & 9.96 \\
    13064645 & -21.00 & -21.29 & 0.75 & 0.88 & 10.63 & 10.96 \\
    12016799 & -20.27 & -19.62 & 1.31 & 0.97 & 10.59 & 9.98 \\
    12023870 & -18.35 & -17.82 & 0.71 & 0.47 & 9.14 & 8.51 \\
    12004470 & -18.40 & -18.64 & 0.64 & 0.48 & 9.07 & 8.86 \\
    12020439 & -18.26 & -18.41 & 0.95 & 0.52 & 9.41 & 8.81 \\

\enddata
\tablecomments{The same twenty randomly-selected galaxies from Table~\ref{tab:1} is shown. The entire catalog is available at: \url{http://people.ucsc.edu/~echeung1/data.html}. 
Col. (1): Unique DEEPID. 
Col. (2): Absolute $B$-band magnitude of bulge derived from Eqn.~\ref{eqn:mb},\ref{eqn:kcorrect}.
Col. (3): Absolute $B$-band magnitude of disk derived from Eqn.~\ref{eqn:mb}-\ref{eqn:kcorrect}.
Col. (4): $U-B$ rest-frame color of bulge derived from Eqn.~\ref{eqn:ub}.
Col. (5): $U-B$ rest-frame color of disk derived from Eqn.~\ref{eqn:ub}.
Col. (6): $M_*$ of bulge derived from Eqn.~\ref{eqn:mb},~\ref{eqn:masslight}.
Col. (7): $M_*$ of disk derived from Eqn.~\ref{eqn:mb},~\ref{eqn:masslight}.
}
\label{tab:2}
\end{deluxetable}

\begin{deluxetable}{lcccccccc}
\tablewidth{0pt} 
\tablecaption{GIM2D: Single $n$ Catalog}
 \tablehead{   
\colhead{DEEPID} &   
\colhead{$V$} &
\colhead{$I$} &
\colhead{$n$}&
\colhead{$e$}&
\colhead{$r_{\rm e}$} &
\colhead{$\chi^2$} &
\colhead{$\chi^2$} 
\\
\colhead{}&
\colhead{ }&  
\colhead{ } &
\colhead{ }&
\colhead{ } &
\colhead{(pixels)} &
\colhead{$V$-band}&
\colhead{$I$-band}& 
\\
\colhead{(1)} & \colhead{(2)} & \colhead {(3)} &
\colhead{(4)} & \colhead{(5)} & \colhead {(6)} &
\colhead{(7)} & \colhead{(8)}
 }
\startdata
       13049654 & 22.07 & 21.56 & 1.11 & 0.68 & 18.31 & 1.16 & 1.07 \\
    13018599 & 24.17 & 22.70 & 1.36 & 0.62 & 15.54 & 1.05 & 1.16 \\
    12020067 & 22.46 & 21.96 & 1.70 & 0.43 & 10.71 & 1.13 & 1.05 \\
    12007757 & 23.73 & 22.84 & 1.17 & 0.58 & 23.76 & 1.09 & 1.03 \\
    13040619 & 22.89 & 21.42 & 6.00 & 0.13 & 19.63 & 1.04 & 0.92 \\
    13048556 & 22.28 & 21.75 & 4.16 & 0.66 & 78.14 & 1.05 & 0.93 \\
    13049852 & 21.80 & 20.41 & 4.79 & 0.23 & 27.88 & 1.00 & 1.12 \\
    13026131 & 23.95 & 23.12 & 1.09 & 0.53 & 15.39 & 0.97 & 0.95 \\
    12020067 & 22.46 & 21.96 & 1.70 & 0.43 & 10.71 & 1.13 & 1.05 \\
    12015606 & 22.48 & 21.54 & 1.59 & 0.03 & 15.41 & 1.08 & 0.98 \\
    13058131 & 22.95 & 21.36 & 5.07 & 0.29 & 34.61 & 1.01 & 0.92 \\
    12016156 & 23.85 & 23.23 & 1.40 & 0.49 & 29.93 & 1.12 & 0.99 \\
    13012297 & 23.31 & 21.67 & 2.82 & 0.15 & 13.04 & 0.89 & 0.82 \\
    12007918 & 22.65 & 21.97 & 1.63 & 0.41 & 10.73 & 1.18 & 1.10 \\
    13011795 & 24.10 & 22.51 & 5.94 & 0.30 & 16.12 & 0.88 & 0.81 \\
    13064645 & 23.97 & 22.49 & 5.99 & 0.25 & 12.85 & 0.96 & 0.93 \\
    12016799 & 23.02 & 21.76 & 1.36 & 0.46 & 21.21 & 1.15 & 1.09 \\
    12023870 & 23.13 & 22.70 & 4.64 & 0.46 & 8.98 & 1.21 & 1.18 \\
    12004470 & 22.71 & 22.34 & 1.66 & 0.41 & 5.67 & 1.62 & 1.48 \\
    12020439 & 23.69 & 23.09 & 1.32 & 0.68 & 6.24 & 1.10 & 1.15 \\
    
   \enddata
\tablecomments{The same twenty randomly-selected galaxies from Table~\ref{tab:1} is shown. The entire catalog is available at: \url{http://people.ucsc.edu/~echeung1/data.html}.
Every value in this table is from the GIM2D decomposition that only fits for a single S\'ersic index. These are the primary data we use throughout this paper, including the basic $V$ and $I$ of the  galaxy. 
Col. (1): DEEPID.
Col. (2): $V$-band magnitude of galaxy. 
Col. (3): $I$-band magnitude of galaxy.
Col. (4): S\'ersic index of galaxy.
Col. (5): Ellipticity of galaxy. $e~\equiv ~1-b/a,~b~\equiv~{\rm semiminor axis},~a~\equiv~{\rm semimajor axis})$
Col. (6): Effective radius of the galaxy measured along the major axis in units of pixels.
Col. (7): $\chi^2$ of fit in the $V$-band.
Col. (8): $\chi^2$ of fit in the $I$-band.
}
\label{tab:3}
\end{deluxetable}

\begin{deluxetable}{lcccccccccccc}
\tablewidth{0pt} 
\tablecaption{GIM2D: $n=4$ Bulge Catalog}
 \tablehead{  
\colhead{DEEPID} &   
\colhead{$V$} &
\colhead{$V$} &
\colhead{$I$} &
\colhead{$I$}&
\colhead{$B/T$}&
\colhead{$B/T$} &
\colhead{$r_{\rm e}$} &
\colhead{$r_{\rm d}$}&
\colhead{$C$} &
\colhead{$C$} &
\colhead{$\chi^2$} &
\colhead{$\chi^2$} \\
\colhead{ }&  
\colhead{Bulge} &
\colhead{Disk} &
\colhead{Bulge} &
\colhead{Disk}&
\colhead{ }&
\colhead{ } &
\colhead{Bulge}&
\colhead{Disk} &
\colhead{ } &
\colhead{ } &
\colhead{ }&
\colhead{ } \\
\colhead{ } &
\colhead{ } &
\colhead{ } &
\colhead{ } &
\colhead{ }&
\colhead{$V$-band}&
\colhead{$I$-band}&
\colhead{(pixels)} &
\colhead{(pixels)} &
\colhead{$V$-band}&
\colhead{$I$-band}&
\colhead{$V$-band}&
\colhead{$I$-band} \\
\colhead{(1)} & \colhead{(2)} & \colhead {(3)} &
\colhead{(4)} & \colhead{(5)} & \colhead {(6)} &
\colhead{(7)} & \colhead{(8)} & \colhead {(9)} &
\colhead{(10)} & \colhead{(11)} & \colhead {(12)} 
& \colhead {(13)}
 }
\startdata
       13049654 & 22.83 & 22.42 & 22.07 & 22.00 & 0.41 & 0.48 & 33.36 & 11.54 & 0.47 & 0.47 & 1.11 & 1.01 \\
    13018599 & 27.39 & 24.32 & 29.81 & 22.78 & 0.06 & 0.00 & 3.27 & 8.56 & 0.48 & 0.58 & 1.08 & 1.20 \\
    12020067 & 23.22 & 23.07 & 22.47 & 22.71 & 0.46 & 0.56 & 19.41 & 5.88 & 0.55 & 0.58 & 1.13 & 1.05 \\
    12007757 & 25.53 & 23.82 & 23.91 & 23.15 & 0.17 & 0.33 & 19.84 & 17.86 & 0.35 & 0.36 & 1.06 & 1.02 \\
    13040619 & 24.11 & 24.14 & 22.87 & 22.28 & 0.51 & 0.37 & 2.78 & 9.96 & 0.62 & 0.69 & 1.01 & 0.89 \\
    13048556 & 23.57 & 23.30 & 23.10 & 22.68 & 0.44 & 0.41 & 23.76 & 29.41 & 0.49 & 0.45 & 1.01 & 0.89 \\
    13049852 & 23.30 & 22.53 & 21.96 & 21.10 & 0.33 & 0.31 & 3.73 & 16.14 & 0.61 & 0.64 & 0.93 & 0.89 \\
    13026131 & 26.47 & 24.05 & 25.78 & 23.21 & 0.10 & 0.09 & 7.93 & 9.66 & 0.41 & 0.46 & 0.97 & 0.96 \\
    12020067 & 23.22 & 23.07 & 22.47 & 22.71 & 0.46 & 0.56 & 19.41 & 5.88 & 0.55 & 0.58 & 1.13 & 1.05 \\
    12015606 & 24.46 & 22.67 & 22.88 & 21.95 & 0.16 & 0.30 & 11.22 & 10.18 & 0.48 & 0.44 & 1.08 & 0.97 \\
    13058131 & 24.43 & 23.65 & 22.59 & 22.42 & 0.33 & 0.46 & 7.95 & 19.11 & 0.57 & 0.55 & 0.97 & 0.88 \\
    12016156 & 28.30 & 23.97 & 25.40 & 23.57 & 0.02 & 0.16 & 18.57 & 15.65 & 0.40 & 0.30 & 1.11 & 1.00 \\
    13012297 & 23.47 & 25.07 & 21.80 & 23.52 & 0.81 & 0.83 & 16.04 & 6.93 & 0.55 & 0.58 & 0.89 & 0.82 \\
    12007918 & 23.45 & 23.19 & 22.35 & 22.74 & 0.44 & 0.59 & 19.23 & 5.87 & 0.55 & 0.56 & 1.16 & 1.07 \\
    13011795 & 24.78 & 25.27 & 23.12 & 24.09 & 0.61 & 0.71 & 6.49 & 12.72 & 0.66 & 0.65 & 0.87 & 0.81 \\
    13064645 & 24.95 & 25.15 & 23.56 & 23.36 & 0.55 & 0.45 & 2.26 & 10.60 & 0.69 & 0.71 & 0.94 & 0.88 \\
    12016799 & 30.56 & 23.11 & 28.18 & 21.85 & 0.00 & 0.00 & 13.39 & 11.62 & 0.36 & 0.38 & 1.17 & 1.14 \\
    12023870 & 23.39 & 24.42 & 22.83 & 24.18 & 0.72 & 0.78 & 21.91 & 1.89 & 0.66 & 0.71 & 1.14 & 1.15 \\
    12004470 & 23.32 & 23.27 & 22.90 & 22.94 & 0.49 & 0.51 & 11.73 & 3.03 & 0.73 & 0.74 & 1.53 & 1.42 \\
    12020439 & 24.77 & 24.01 & 23.65 & 23.61 & 0.33 & 0.49 & 13.89 & 3.54 & 0.65 & 0.72 & 1.02 & 1.06 \\
  \enddata
\tablecomments{The same twenty randomly-selected galaxies from Table~\ref{tab:1} is shown. The entire catalog is available at: \url{http://people.ucsc.edu/~echeung1/data.html}.
Every value in this table is from the GIM2D decomposition using a bulge and disk with S\'ersic index of $n=4$ and $n=1$, respectively.
Col. (1): DEEPID.
Col. (2) \& (3): $V$-band magnitude of bulge and disk.
Col. (4) \& (5): $I$-band magnitude of bulge and disk.
Col. (6): $B/T$ in $V$-band.
Col. (7): $B/T$ in $I$-band.
Col. (8): Effective radius of major axis of bulge in units of pixels.
Col. (9): Scale length of disk measured along major axis in units of pixel.
Col. (10): Concentration with $\alpha=0.3$ in $V$.
Col. (11): Concentration with $\alpha=0.3$ in $I$.
Col. (12): Chi-squared of fit in $V$.
Col. (13): Chi-squared of fit in $I$.
}
\label{tab:4}
\end{deluxetable}

\begin{deluxetable}{lcccccccccccc}
\tablewidth{0pt} 
\tablecaption{GIM2D: $n=2$ Bulge Catalog}
 \tablehead{   
\colhead{DEEPID} &   
\colhead{$V$} &
   \colhead{$I$} &
   \colhead{$V$} &
\colhead{$I$}&
\colhead{$B/T$}&
\colhead{$B/T$} &
\colhead{$r_{\rm e}$} &
\colhead{$r_{\rm e}$}&
\colhead{$C$} &
\colhead{$C$} &
\colhead{$\chi^2$} &
\colhead{$\chi^2$} \\
\colhead{ }&
  \colhead{Bulge} &
   \colhead{Bulge} &
   \colhead{Disk} &
\colhead{Disk}&
\colhead{ }&
\colhead{ } &
\colhead{Bulge}&
\colhead{Disk} &
\colhead{ } &
\colhead{ } &
\colhead{ }&
\colhead{ } \\
\colhead{ } &
\colhead{ } &
\colhead{ } &
\colhead{ } &
\colhead{ }&
\colhead{$V$-band}&
\colhead{$I$-band}&
\colhead{(pixels)} &
\colhead{(pixels)} &
\colhead{$V$-band}&
\colhead{$I$-band}&
\colhead{$V$-band}&
\colhead{$I$-band} \\
\colhead{(1)} & \colhead{(2)} & \colhead {(3)} &
\colhead{(4)} & \colhead{(5)} & \colhead {(6)} &
\colhead{(7)} & \colhead{(8)} & \colhead {(9)} &
\colhead{(10)} & \colhead{(11)} & \colhead {(12)} 
& \colhead {(13)}
}  
\startdata
     13049654 & 23.33 & 22.38 & 22.51 & 21.90 & 0.29 & 0.36 & 24.61 & 10.97 & 0.47 & 0.47 & 1.11 & 1.00 \\
    13018599 & 25.53 & 24.55 & 23.41 & 23.23 & 0.29 & 0.46 & 20.65 & 8.46 & 0.48 & 0.58 & 1.05 & 1.12 \\
    12020067 & 23.02 & 23.16 & 22.43 & 22.70 & 0.53 & 0.56 & 23.38 & 4.43 & 0.53 & 0.55 & 1.05 & 0.98 \\
    12007757 & 24.66 & 23.79 & 23.48 & 23.18 & 0.31 & 0.43 & 14.76 & 35.53 & 0.35 & 0.36 & 1.04 & 0.98 \\
    13040619 & 24.40 & 23.95 & 23.20 & 22.18 & 0.40 & 0.28 & 1.92 & 9.09 & 0.62 & 0.69 & 1.01 & 0.90 \\
    13048556 & 22.99 & 25.33 & 22.59 & 23.80 & 0.90 & 0.75 & 28.14 & 30.87 & 0.50 & 0.48 & 1.11 & 0.96 \\
    13049852 & 23.48 & 22.48 & 22.14 & 21.05 & 0.28 & 0.27 & 3.13 & 15.79 & 0.61 & 0.64 & 0.93 & 0.88 \\
    13026131 & 24.93 & 24.34 & 23.73 & 23.68 & 0.37 & 0.49 & 19.07 & 9.98 & 0.41 & 0.46 & 0.95 & 0.93 \\
    12020067 & 23.02 & 23.16 & 22.43 & 22.70 & 0.53 & 0.56 & 23.38 & 4.43 & 0.53 & 0.55 & 1.05 & 0.98 \\
    12015606 & 23.71 & 22.89 & 22.36 & 22.30 & 0.32 & 0.49 & 10.64 & 11.16 & 0.48 & 0.44 & 1.07 & 0.97 \\
    13058131 & 24.81 & 23.55 & 22.95 & 22.23 & 0.24 & 0.34 & 4.94 & 17.80 & 0.57 & 0.55 & 0.97 & 0.88 \\
    12016156 & 24.67 & 24.31 & 23.93 & 23.91 & 0.42 & 0.49 & 27.86 & 30.96 & 0.40 & 0.30 & 1.11 & 0.96 \\
    13012297 & 23.44 & -99.99 & 21.80 & -99.99 & 1.00 & 1.00 & 10.84 & 6.88 & 0.55 & 0.58 & 0.91 & 0.88 \\
    12007918 & 23.33 & 23.54 & 22.77 & 22.68 & 0.55 & 0.48 & 9.21 & 7.01 & 0.55 & 0.56 & 1.17 & 1.08 \\
    13011795 & 25.30 & 24.90 & 23.63 & 23.50 & 0.41 & 0.47 & 3.11 & 13.02 & 0.66 & 0.65 & 0.87 & 0.80 \\
    13064645 & 25.29 & 24.90 & 23.93 & 23.21 & 0.41 & 0.34 & 1.36 & 8.69 & 0.69 & 0.71 & 0.94 & 0.89 \\
    12016799 & 23.41 & 23.85 & 21.90 & 22.78 & 0.60 & 0.69 & 44.45 & 9.40 & 0.36 & 0.38 & 1.11 & 1.06 \\
    12023870 & 23.71 & 24.13 & 23.12 & 23.82 & 0.60 & 0.66 & 21.50 & 1.75 & 0.66 & 0.71 & 1.14 & 1.15 \\
    12004470 & 23.50 & 23.19 & 23.00 & 22.89 & 0.43 & 0.47 & 14.01 & 2.61 & 0.73 & 0.74 & 1.48 & 1.37 \\
    12020439 & 24.56 & 24.16 & 23.63 & 23.72 & 0.41 & 0.52 & 12.21 & 3.11 & 0.65 & 0.72 & 0.98 & 1.05 \\

 \enddata
\tablecomments{The same twenty randomly-selected galaxies from Table~\ref{tab:1} is shown. The entire catalog is available at: \url{http://people.ucsc.edu/~echeung1/data.html}.
Every value in this table is from the GIM2D decomposition using a bulge and disk with S\'ersic index of $n=2$ and $n=1$, respectively.
Col. (1): DEEPID.
Col. (2) \& (3): $V$-band magnitude of bulge and disk.
Col. (4) \& (5): $I$-band magnitude of bulge and disk.
Col. (6): $B/T$ in $V$-band.
Col. (7): $B/T$ in $I$-band.
Col. (8): Effective radius of major axis of bulge in units of pixels.
Col. (9): Scale length of disk measured along major axis in units of pixel.
Col. (10): Concentration with $\alpha=0.3$ in $V$.
Col. (11): Concentration with $\alpha=0.3$ in $I$.
Col. (12): Chi-squared of fit in $V$.
Col. (13): Chi-squared of fit in $I$.
}
\label{tab:5}
\end{deluxetable}

\end{document}